\documentclass[12pt]{iopart}
\usepackage{amssymb,graphicx,iopams,setstack}

\newcommand{\ket}[1]{|#1\rangle}
\newcommand{\bra}[1]{\langle#1|}

\newcommand{\zb}{{\bar{z}}}
\newcommand{\vb}{{\bar{v}}}
\newcommand{\zzbar}{(z,\bar{z})}
\newcommand{\Jp}{\hat{J}_+}

\newcommand{\matel}[3]{\langle #1|#2|#3\rangle}
\newcommand{\inp}[2]{\langle #1|#2\rangle}

\newcommand{\Ts}{\mathcal{T}}
\newcommand{\Is}{\mathcal{I}}
\newcommand{\Ds}{\mathcal{D}}
\newcommand{\Fs}{\mathcal{F}}
\newcommand{\Ss}{\mathcal{S}}
\newcommand{\oforder}[1]{\mathcal{O}(#1)}
\newcommand{\dx}{{\rm d}x\,}
\newcommand{\dt}{{\rm d}\theta\,}
\newcommand{\inflim}[1]{\lim_{#1\rightarrow\infty}}

\newcommand{\zeroplim}[1]{\lim_{#1\rightarrow0^+}}
\newcommand{\hs}[1]{\hspace{#1 cm}}
\newcommand{\az}{\alpha_0}
\newcommand{\bz}{\beta_0}

\newcommand{\binom}[2]{{#1 \choose #2}} 
\newcommand{\mt}[2]{\hs{#2}\mbox{#1}\hs{#2}}

\def\imagetop#1{\vtop{\null\hbox{#1}}}
  
\begin{document}
\title{Eigenvalue distributions from a star product approach}  
\date{\today}
\author{J N Kriel$^1$ and F G Scholtz$^{1,2}$}
\address{$^1$ Institute of Theoretical Physics, University of Stellenbosch, Stellenbosch 7600, South Africa}
\address{$^2$ National Institute for Theoretical Physics (NITheP), 7600 Stellenbosch, South Africa}

\begin{abstract}
We use the well-known isomorphism between operator algebras and function spaces equipped with a star product to study the asymptotic properties of certain matrix sequences in which the matrix dimension $D$ tends to infinity. Our approach is based on the $su(2)$ coherent states which allow for a systematic $1/D$ expansion of the star product. This produces a trace formula for functions of the matrix sequence elements in the large-$D$ limit which includes higher order (finite-$D$) corrections. From this a variety of analytic results pertaining to the asymptotic properties of the density of states, eigenstates and expectation values associated with the matrix sequence follows. It is shown how new and existing results in the settings of collective spin systems and orthogonal polynomial sequences can be readily obtained as special cases.  In particular, this approach allows for the  calculation of higher order corrections to the zero distributions of a large class of orthogonal polynomials.  
\end{abstract}

\section{Introduction}
The eigenvalue analysis of operators that arise in physical problems is often simplified by exploiting special structures present in the operator's matrix representation. One such structure encountered in some matrix sequences $\{M_D\}_{D=1}^{\infty}$, where $M_D$ is $D\times D$ dimensional, concerns the limiting behaviour of the matrix elements $[M_{D}]_{n,n+m}$ as $n$ and $D$ tend to infinity in a fixed ratio. In particular, there may exist a set of functions $\{f_{m}(x)\}_{m=0}$ such that $[M_D]_{n,n+m}\rightarrow f_m(x)$ when $n,D\rightarrow\infty$ and $n/D\rightarrow x$. Intuitively, this suggests that the matrix elements on the $m$'th diagonal represent an increasingly fine sampling of $f_m(x)$ at the points $x_n=n/D$. If each $f_m(x)$ is continuous the matrix elements therefore exhibit a ``smooth" structure in that $[M_{D}]_{n,n+m}$ is a slowly varying function of $n$ when $D$ is large. 
Since the set $\{f_m(x)\}$ characterises the matrix sequence as $D\rightarrow\infty$ one expects that the asymptotic behaviour of the eigenvalue distribution, eigenstates and expectation values can be studied directly in terms of these functions. The development of such a formalism is the main goal of this paper.\\

Despite their usefulness these ideas appear to have found only limited application in physical problems. Hollenberg and Witte \cite{hollenberg1996} made use of this structure in the Hamiltonian matrices of extensive many-body systems to derive an exact analytic expression for the ground state energy density by minimizing a particular linear combination of the $f_m(x)$'s.  Deift and Mclaughlin \cite{deift1998} (see also \cite{aptekarev2001}) studied a continuum limit of the Toda lattice system of which the equations of motion amount to the isospectral flow of a particular tridiagonal matrix with a dimension given by the system size. Under the conditions outlined above the continuum limit of this matrix equation is a partial differential equation for the $f_m(x)$'s. In fact, from the matrix viewpoint the Toda lattice dynamics represent a special case of the flow equation renormalization scheme proposed by Wegner \cite{wegner1994} which has been applied to a wide range of quantum mechanical many-body problems \cite{kehrein2006}. Here too the presence of a smooth structure in the matrix elements allows the matrix (or operator) flow to be recast as a partial differential equation in the thermodynamic limit \cite{kriel2005,kriel2007}. An important feature of these studies is their non-perturbative nature, i.e. no assumptions regarding the magnitude of coupling constants are made and expansions are instead controlled by the small parameter $1/D$. These techniques are therefore particularly well suited to the study of quantum critical behaviour in the thermodynamic limit \cite{kriel2007}.\\

Matrix sequences of this type also arise naturally in the theory of orthogonal polynomials. Kuijlaars and van Assche \cite{kuijlaars1999,kuijlaars2001} made use of this structure in the Jacobi matrix to derive an expression for the asymptotic zero distribution of certain orthogonal polynomial sequences. See also Bourget \cite{bourget2011} for a recent application to the theory of Jacobi matrices of which the elements satisfy a similar small deviation condition.\\

In this paper we present a treatment of matrix sequences which exhibit this smooth structure using a coherent state star-product formalism. The advantage of this approach is that it produces, in a simple and direct manner, analytic results for the asymptotic eigenvalue distribution (i.e. the density of states) as well as for eigenstates and expectation values. Indeed, our approach is based on a systematic $1/D$-expansion which allows for the inclusion of finite-$D$ corrections. This is in contrast to most existing results which only consider the $D\rightarrow\infty$ limit. In particular we are able to extend the results of \cite{kuijlaars1999} and derive corrections to the asymptotic zero distributions of a large class of orthogonal polynomial sequences. In quantum mechanical applications this allows for the inclusion of finite-size corrections in the density of states and expectation values.\\

The paper is organized as follows. In section \ref{sectionbackground} we summarise the basic properties of spin coherent states and introduce the notions of symbols and the star-product. The precise class of real tridiagonal matrix sequences under consideration is defined in section \ref{sectionsmoothsymbols} and asymptotic expressions for their symbols are derived. Section \ref{sectiontraces} is dedicated to the derivation of a trace formula which is used in section \ref{sectionDOS} to derive our main result: an expression for the asymptotic eigenvalue distribution which includes higher order corrections. Section \ref{sectionspecialcases} deals with expectation values and the structure of eigenstates. Results obtained for tridiagonal sequences can be generalized to sequences of banded Hamiltonian matrices satisfying certain boundary conditions. We show how this is done in section \ref{sectionbanddiagonal}. Sections \ref{sectioncollective} and \ref{sectionortpol} present applications of our results to collective quantum spin systems and orthogonal polynomials.

\section{Background}
\label{sectionbackground}
\subsection{Spin coherent states}
\label{sectioncs}
Let $\mathcal{H}_j$ be a $D=2j+1$ dimensional Hilbert space carrying the $j$-irrep of $su(2)$ with generators $\{\hat{J}_\pm,\hat{J}_z\}$. The spin coherent states \cite{perelomov1986, klauder1985} are defined as
\begin{equation}
	\ket{z}=(1+z\zb)^{-j}\exp[z\Jp]\ket{j,-j}\hs{1.5}z\in\mathbb{C}.
\end{equation}
Here $\zb$ is the conjugate of $z$ and $\{\ket{j,m}\}_{m=-j}^j$ is the standard basis of $\hat{J}_z$ eigenstates. The inner product of any two such states is
\begin{equation}
	|\inp{z}{v}|^2=(1+z\vb)^{2j}(1+v\zb)^{2j}(1+z\zb)^{-2j}(1+v\vb)^{-2j}.
\end{equation}
The set of coherent states forms an over-complete basis for $\mathcal{H}_j$ and provides a resolution of the identity in the form $\hat{I}=\int{\rm d}z\,{\rm d}\zb\,\mu\zzbar\ket{z}\bra{z}$ with \mbox{$\mu\zzbar=(2j+1)\pi^{-1}(1+z\zb)^{-2}$}.\\

Let $z=r\exp(i\theta)$ and set $x=r^2/(1+r^2)\in[0,1]$. The expansion coefficients of $\ket{z}$ in the $\hat{J}_z$-basis are 
\begin{equation}
	\fl c_n=\inp{j,-j+n}{z}=x^{n/2} (1-x)^{(2j-n)/2} e^{in\theta}\binom{2j}{n}^{1/2}\hs{1.5}n=0,1,2,\ldots,2j.
\end{equation}
Note that $P(n;2j,x)=|c_n|^2$ amounts to a binomial distribution with success probability $x$ and $2j$ trails. It is known that if $x\in(0,1)$ is kept fixed while $j$ tends to infinity then $P(n;2j,x)$ will approach a normal distribution with mean $2jx$ and variance $2jx(1-x)$. 
\subsection{Symbols and the star product}
\label{sectionstarproduct}
With each operator $\hat{H}$ acting on $\mathcal{H}_j$ we associate a function $H\zzbar=\matel{z}{\hat{H}}{z}$ known as the symbol of $\hat{H}$. The coherent state resolution of the identity allows the trace of $\hat{H}$ to be expressed in terms of its symbol as
\begin{equation}
	{\rm tr}(\hat{H})=\int{\rm d}z\,{\rm d}\zb\,\mu\zzbar H\zzbar=\frac{2j+1}{2 \pi}\int_0^1{\rm d}x\int_0^{2\pi}{\rm d}\theta\,H(x,\theta).
	\label{traceasintegral}
\end{equation}
In what follows we use the $\zzbar$ and $(x,\theta)$ parametrizations interchangeably and will often suppress the functional dependence of the symbols.\\

The operator product is realised on the symbol level through the star product
\begin{equation}
	\matel{z}{\hat{A}\hat{B}}{z}=\matel{z}{\hat{A}}{z}\ast\matel{z}{\hat{B}}{z}=A\ast B
	\label{fullstar}
\end{equation}
which can be expressed in terms of differential operators acting on the symbols as \cite{alexanian2001}
\begin{equation}
	\ast=\int{\rm d}v\,{\rm d}\vb\,\exp[v\overleftarrow{\partial}_{z}]\,\mu(v+z,\vb+\zb)|\inp{z}{v+z}|^2\, \exp[\vb\overrightarrow{\partial}_\zb].
	\label{starintegral}
\end{equation}
This exact representation can be brought into a more practical form by noting that $|\inp{z}{v+z}|^2$ is maximal at $v=0$ and that expanding its logarithm around this point produces, to lowest order, the quadratic expression 
\begin{equation}
	\log|\inp{z}{v+z}|^2=\frac{-2j}{(1+z\zb)^2}v\vb+\cdots
\end{equation}
This suggests that a saddle-point approximation of the integral in \eref{starintegral} should allow the star product to be expressed as a power series in $1/j$.
The result can be written as
\begin{equation}
\ast=1+\sum_{k=1}^\infty\frac{1}{j^k}\sum_{n,m=1}^k\overleftarrow{\partial}^n_z(1+z\zb)^{n+m}\Lambda^{(k)}_{n,m}\overrightarrow{\partial}^m_\zb
\label{starexpand}
\end{equation}
where $\Lambda^{(k)}$ is a $k\times k$ Hermitian matrix which generally contains $z$ and $\zb$. The first three of these are $\Lambda^{(1)}=1/2$, 
\begin{equation}
	\fl \Lambda^{(2)}=\frac{1}{8}\left[ \begin{array}{cc} 4 z\zb & 2\zb \\ 2 z & 1 \end{array} \right]\hs{0.8}{\rm and}\hs{0.8}\Lambda^{(3)}=\frac{1}{48}\left[\begin{array}{ccc} 12 z\zb (3 z\zb+1) & 6\zb (6 z\zb+1) & 6 \zb^2 \\ 6 z (6 z\zb+1) & 36 z\zb+3 & 6\zb \\ 6 z^2 & 6 z & 1 \end{array}\right].
	\label{starexpandmatrices}
\end{equation}
To linear order in $1/j$ the star product is therefore given by
\begin{equation}
	\ast_L\equiv1+\overleftarrow{\partial}_z\frac{(1+z\zb)^2}{2j}\overrightarrow{\partial}_\zb.
	\label{truncstar}
\end{equation}
\subsection{Smooth symbols and fluctuations}
\label{sectionsmoothness}
Consider a sequence of operators $\{\hat{A}_j\}_{2j=0}^\infty$ where $\hat{A}_j$ acts on the $D=2j+1$ dimensional space $\mathcal{H}_j$. We assume that the corresponding sequence of symbols $\{A_j\}$ \emph{as well as their derivatives} to $z$ and $\zb$ scale intensively with $j$, i.e. as $\oforder{j^0}$. This smoothness condition allows the symbol $\matel{z}{\hat{A}_j^2}{z}=A_j\ast A_j$ to be approximated to linear order in $1/j$ as
\begin{equation}
	A_j \ast_L A_j=A_j^2+\frac{(1+z\zb)^2}{2j}\left(\partial_z A_j\right)\left(\partial_\zb A_j\right).
	\label{producttrunc}
\end{equation}
As shown next this result has important consequences for the scaling behaviour of fluctuations. Since the symbol $A_j$ is an expectation value with respect to $\ket{z}$ we write $\delta\hat{A}_j=\hat{A}_j-A_j$, in which case $\matel{z}{(\delta\hat{A}_j)^2}{z}=(1+z\zb)^2\left(\partial_z A_j\right)\left(\partial_\zb A_j\right)/(2j)+\mathcal{O}(j^{-2})$. The fluctuations in $\hat{A}_j$ with respect to the coherent states are therefore suppressed by a factor of $1/j$. This result may well have been anticipated due to the semi-classical nature of these states, but here we wish to emphasise the importance of the smoothness condition. It follows from induction that higher-order fluctuations are even further suppressed, with $\matel{z}{(\delta\hat{A}_j)^n}{z}$ generally scaling like $\mathcal{O}(j^{-\lceil n/2\rceil})$.\\

We arrive at the useful result that for a smooth function $f(x)$ the symbol of $f(\hat{A})$ is given, up to linear order in $1/j$, by
\begin{equation}
	\matel{z}{f(\hat{A}_j)}{z}=f(A_j)+\frac{(1+z\zb)^2}{4j}f''(A_j)\left(\partial_z A_j\right)\left(\partial_\zb A_j\right)+\mathcal{O}(j^{-2}).
	\label{genfexpand}
\end{equation}
This simple expression forms the cornerstone of the asymptotic expansions that follow.
\section{Matrices with smoothly varying elements}
\label{sectionsmoothsymbols}
Next we define the class of matrix sequences to be considered in the sequel. For simpli-city we will focus mainly on the tridiagonal case, although all the results can be adapted to apply to band diagonal Hermitian matrices of which the off-diagonal elements satisfy a particular constraint. This straightforward generalization is presented in section \ref{sectionbanddiagonal}.\\

Let $\{\hat{H}_j\}$ denote a sequence of real, symmetric tridiagonal matrices with $\hat{H}_j$ acting on the $D=2j+1$ dimensional space $\mathcal{H}_j$. The basis in which $\hat{H}_j$ is given is identified with the basis $\{\ket{j,m}\}_{m=-j}^j$ of $\hat{J}_z$ eigenstates. The diagonal matrix elements of $\hat{H}_j$ are denoted by $a_{n,j}=[\hat{H}_j]_{n+1,n+1}$ with $n=0,\ldots,2j$ while the off-diagonal ones are $b_{n,j}=[\hat{H}_j]_{n+1,n}$ with $n=1,\ldots,2j$. We require that these matrix elements satisfy a certain smoothness condition in that they are slowly varying functions of $n$. To be precise, we assume that there exists smooth functions $A_0,A_1,B_0,B_1:[0,1]\rightarrow\mathbb{R}$ such that, up to linear order in $1/j$, the matrix elements are given by
\begin{equation}
	\fl a_{n,j}=A_0\left(\frac{n}{2j}\right)+\frac{1}{j}A_1\left(\frac{n}{2j}\right)\ \ \ \ \ {\rm and}\ \ \ \ \ b_{n,j}=B_0\left(\frac{2n-1}{4j}\right)+\frac{1}{j}B_1\left(\frac{2n-1}{4j}\right).
	\label{ABdef}
\end{equation}
where $B_0(x)>0$ for $x\in(0,1)$. Note that $\hat{H}_j$ scales intensively with $j$.\\

Our first task is to find a large-$j$ expansion for the symbol $H_j$ of $\hat{H}_j$. Some algebra reveals that $H_j$ is given exactly by
\begin{equation}
	\fl \matel{z}{\hat{H}_j}{z}=H_j(x,\theta)=\left\langle a_{n,j}\right\rangle_x+2 \left[\frac{x}{1-x}\right]^{1/2}\left\langle\left[\frac{2j-n}{n+1}\right]^{1/2}b_{n+1,j}\right\rangle_x\cos(\theta)
	\label{Hsymbol0}
\end{equation}
where $\langle f(n)\rangle_x$ denotes the expectation value of $f(n)$ over $n$ with respect to the binomial distribution $P(n;2j,x)=|c_n|^2$. Now consider a fixed $x\in(0,1)$. At large $j$ the distribution $P(n;2j,x)$ is sharply peaked around the mean $\langle n \rangle=2jx$ with fluctuations in $n/2j$ of order $\oforder{j^{-1/2}}$. This suggests that the expectation values in \eref{Hsymbol0} can be approximated using an expansion in orders of $\delta n=n-\langle n\rangle$. Up to linear order in $1/j$ and for $x\in(0,1)$ this produces 
\begin{equation}
	H_j(x,\theta)=H^{(0)}(x,\theta)+H^{(1)}(x,\theta)/j+\oforder{j^{-2}}
\end{equation}
where 
\begin{eqnarray}
	\fl H^{(0)}(x,\theta)&=&A_0(x)+2 B_0(x)\cos(\theta)\nonumber\\
	\fl H^{(1)}(x,\theta)&=&A_1(x)+2 B_1(x)\cos(\theta)+\frac{(1-x)x}{4}\,\partial_x^2H^{(0)}(x,\theta)-\frac{B_0(x) \cos(\theta)}{8(1-x)x}.\hs{1}
	\label{H0H1general}
\end{eqnarray}
Note that the functions $H^{(0)}(x,\theta)$ and $H^{(1)}(x,\theta)$ can also be defined for $x\in[0,1]$ through the limits
\begin{equation}
	\fl H^{(0)}(x,\theta)=\lim_{j\rightarrow\infty}H_j(x,\theta)\ \ \ \ {\rm and}\ \ \ \ 	H^{(1)}(x,\theta)=\lim_{j\rightarrow\infty} j(H_j(x,\theta)-H^{(0)}(x,\theta)).
\end{equation}
A natural question is whether the results of \eref{H0H1general}, which were valid for $x\in(0,1)$, also hold at $x=0,1$. We will show that this is not generally the case. That the end-points require special attention is a consequence of the fundamentally different nature of the coherent state $\ket{z}$ for $x=0,1$ compared to $x\in(0,1)$. When $x\in(0,1)$ the state $\ket{z}$ is a linear combination of \emph{all} $2j+1$ basis states and the symbol $H_j=\matel{z}{\hat{H}_j}{z}$ therefore depends on all the matrix elements of $\hat{H}_j$. In contrast, at $x=0,1$ the state $\ket{z}$ is a single basis state, either $\ket{j,-j}$ or $\ket{j,+j}$, and $\matel{z}{\hat{H}_j}{z}$ is simply the top left or bottom right diagonal matrix element. It follows that
\begin{eqnarray}
	H^{(0)}(0,\theta)&=&A_0(0)\hs{2}H^{(0)}(1,\theta)=A_0(1)\nonumber\\
	H^{(1)}(0,\theta)&=&A_1(0)\hs{2}H^{(1)}(1,\theta)=A_1(1)
\end{eqnarray}
which is to be compared with the results for $x\in(0,1)$ in \eref{H0H1general}. The continuity of $H^{(0,1)}(x,\theta)$ at $x=0,1$ clearly depends on the behaviour of the functions $B_0(x)$ and $B_1(x)$ at the edges. In particular, continuity to lowest order requires that $B_0(x)$ vanishes at $x=0,1$. We say that a matrix sequence is \emph{closed} if $B_0(0)=B_0(1)=0$. This implies that the off-diagonal matrix elements approach zero at the top left and bottom right corners of the matrix as $j\rightarrow\infty$. Continuity in the higher order terms will clearly place further restrictions on the behaviour of $B_0(x)$ and $B_1(x)$ at $x=0,1$. For our purposes the notion of closure at lowest order in $1/j$ is sufficient.\\
\begin{figure}[t]
\begin{tabular}{cc}
\includegraphics[width=7.7cm]{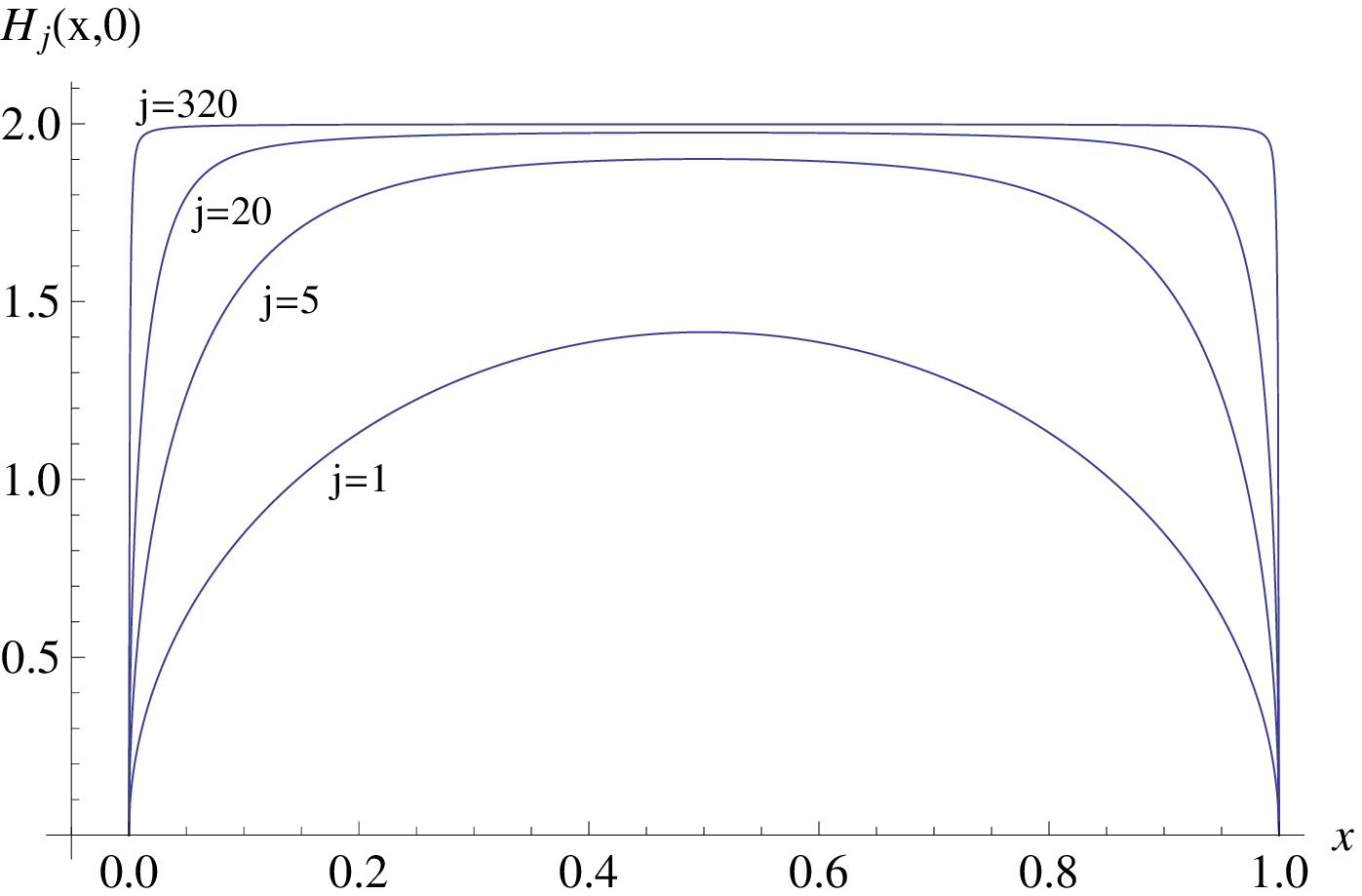} & \includegraphics[width=7.7cm]{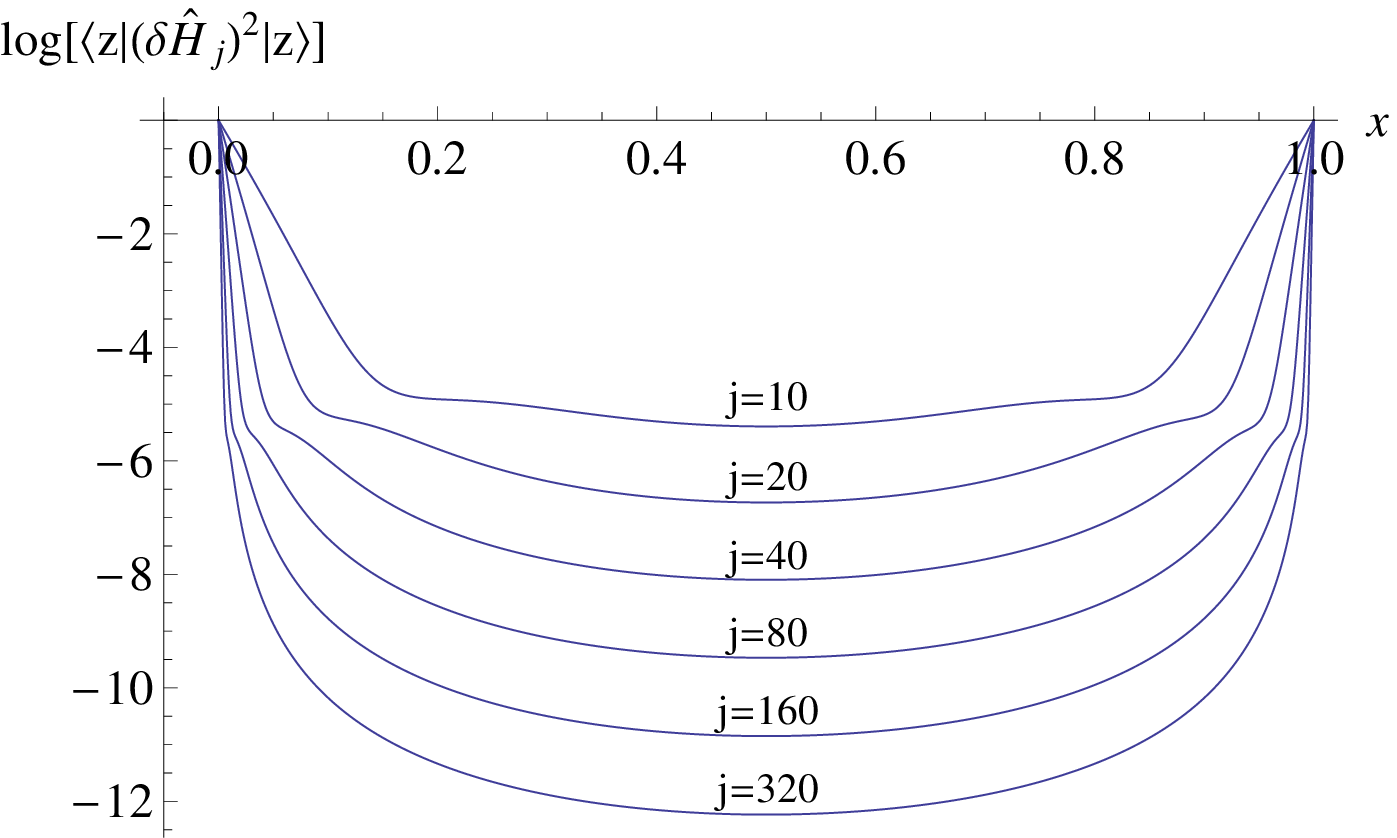}\\
(a) & (b)
\end{tabular}
\caption{Results for the Toeplitz sequence with $a=0$ and $b=1$ at finite $j$. (a) The symbol $H_j(x,\theta)$ as a function of $x$ at $\theta=0$. (b) The logarithm of $\matel{z}{(\delta\hat{H}_j)^2}{z}$ for $z=\sqrt{x/(1-x)}$.}
\label{toeplitzplots}
\end{figure}
To illustrate how the continuity of the symbols depend on the behaviour of $B_0(x)$ at $x=0,1$ we consider as an example a sequence of Toeplitz matrices. Since these issues are local in nature what is observed here is quite generic and common to all non-closed matrix sequences. For tridiagonal Toeplitz matrices with constant entries $a_{n,j}=a$ and $b_{n,j}=b>0$ we have $H^{(0)}(0,\theta)=H^{(0)}(1,\theta)=a$ while $H^{(0)}(x,\theta)=a+2b\cos(\theta)$ when $x\in(0,1)$. The sequence of continuous symbols $H_j$ therefore converge to a function which is continuous everywhere except at $x=0,1$. This is reflected by the numerical results in figure \ref{toeplitzplots} (a). Furthermore, in a neighbourhood of $x=0,1$ the symbols $H_j$ clearly vary very rapidly and have derivatives which scale like $j$, despite the symbols themselves being intensive. From the discussion in section \ref{sectionsmoothness} this is expected to impact on the scaling behaviour of the fluctuations close to the edges. This is indeed the case, as can be seen in figure \ref{toeplitzplots} (b). The expansion in \eref{genfexpand} for $\matel{z}{f(\hat{H}_j)}{z}$ therefore fails at $x=0,1$. 

\section{Asymptotic expansion of traces}
\label{sectiontraces}
Let $\{\hat{H}_j\}$ be a matrix sequence as defined section \ref{sectionsmoothsymbols} and $f(x)$ a smooth bounded function. We are interested in the large-$j$ behaviour of the scaled trace
\begin{equation}
 \Ts_j=\frac{1}{2j+1}\tr(f(\hat{H}_j))=\frac{1}{2\pi}\int_0^1\dx\int_0^{2\pi}\dt\matel{z}{f(\hat{H}_j)}{z} 
 \label{tracedef}
\end{equation}
which is captured by the expansion
\begin{equation}
	\Ts_j=\Ts^{(0)}+\Ts^{(1)}/j+\oforder{j^{-2}}
	\label{Tsexpand}
\end{equation}
where 
\begin{equation}
	\Ts^{(0)}=\lim_{j\rightarrow\infty}\Ts_j\hs{1.2}{\rm and}\hs{1.2}\Ts^{(1)}=\lim_{j\rightarrow\infty} j(\Ts_j-\Ts^{(0)}).
	\label{Tlimits}
\end{equation}
For $x\in(0,1)$ the expansion in \eref{genfexpand} produces 
\begin{equation}
	\fl \matel{z}{f(\hat{H}_j)}{z}=f(H^{(0)})+\frac{1}{j}\left[f'(H^{(0)}) H^{(1)}+\frac{1}{4}f''(H^{(0)})(1+z\zb)^2\left(\partial_z H^{(0)}\right)\left(\partial_\zb H^{(0)}\right)\right].
	\label{fofHexpansion}
\end{equation}
with $H^{(0)}$ and $H^{(1)}$ given by \eref{H0H1general}. We will show that for closed sequences this result suffices to calculate both $\Ts^{(0)}$ and $\Ts^{(1)}$. However, for sequences which are not closed the uncontrolled fluctuations at $x=0,1$ result in non-trivial edge corrections to $\Ts^{(1)}$ which are not captured by the expansion above. 
\subsection{The lowest order contribution $\Ts^{(0)}$}
To determine $\Ts^{(0)}$ we first split the integral in \eref{tracedef} into three parts to isolate the problematic edges at $x=0,1$ from the interior where \eref{fofHexpansion} holds:
\begin{eqnarray}
\fl \Ts^{(0)}&=&\inflim{j}\int_0^1\dx\int_0^{2\pi}\frac{\dt}{2\pi}\matel{z}{f(\hat{H}_j)}{z}\\
\fl &=&\zeroplim{\epsilon}\inflim{j}\left[\int_0^\epsilon\dx(\ldots)+\int_\epsilon^{1-\epsilon}\dx(\ldots)+\int_{1-\epsilon}^1\dx(\ldots)\right]\\
\fl &=&\underbrace{\zeroplim{\epsilon}\inflim{j}\int_0^\epsilon\dx(\ldots)}_{\Ts^{(0)}_{L}}+\underbrace{\zeroplim{\epsilon}\inflim{j}\int_\epsilon^{1-\epsilon}\dx(\ldots)}_{\Ts^{(0)}_I}+\underbrace{\zeroplim{\epsilon}\inflim{j}\int_{1-\epsilon}^1\dx(\ldots)}_{\Ts^{(0)}_{R}}\hs{0.5}
 \label{T0finalline}
\end{eqnarray}
Here $\Ts^{(0)}_I$, $\Ts^{(0)}_L$ and $\Ts^{(0)}_R$ denote the contributions to $\Ts^{(0)}$ coming from the interior and left and right edges of the $x\in[0,1]$ interval respectively. Since $f$ is bounded the two edge terms are zero. From the expressions for $H^{(0)}$ and $H^{(1)}$ in \eref{H0H1general} we also see that the $\oforder{j^{-1}}$ terms in \eref{fofHexpansion} can only diverge (if at all) at $x=0,1$ and may therefore be neglected when taking the $j\rightarrow\infty$ limit of $\int_\epsilon^{1-\epsilon}\dx(\ldots)$ for any finite $\epsilon$. It follows that
\begin{equation}
	\fl \Ts^{(0)}=\int_0^1\dx\int_0^{2\pi}\frac{\dt}{2\pi}f(H^{(0)}(x,\theta))=\int_0^1\dx\int_0^{2\pi}\frac{\dt}{2\pi}f(A_0(x)+2 B_0(x)\cos(\theta)).
	\label{T0general}
\end{equation}
To lowest order in $1/j$ the edge effects are therefore irrelevant and \eref{T0general} above holds regardless of whether or not the matrix sequence is closed. 
\subsection{The linear order correction $\Ts^{(1)}$}
From \eref{Tlimits} and \eref{T0general} the linear order correction to $\Ts_j$ can be written as
\begin{equation}
	\Ts^{(1)}=\inflim{j}\int_0^1\dx\int_0^{2\pi}\frac{\dt}{2\pi} j\left[\matel{z}{f(\hat{H}_j)}{z}-f(H^{(0)})\right].
	\label{T1Integral}
\end{equation}
Here we again isolate the edge corrections by writing $\Ts^{(1)}=\Ts^{(1)}_{L}+\Ts^{(1)}_{I}+\Ts^{(1)}_{R}$ as was done for $\Ts^{(0)}$ in \eref{T0finalline}. We consider these three contributions individually.
\subsubsection{The interior term $\Ts^{(1)}_{I}$}
\label{subsectioninterior}
Since the integral for $\Ts^{(1)}_{I}$ excludes the edges it follows from \eref{fofHexpansion} that
\begin{equation}
	\fl \Ts^{(1)}_{I}=\zeroplim{\epsilon}\int_\epsilon^{1-\epsilon}\hs{-0.4}\dx\int_0^{2\pi}\frac{\dt}{2\pi}\left[f'(H^{(0)}) H^{(1)}+\frac{1}{2}f''(H^{(0)})\left[\frac{(\partial_\theta H^{(0)})^2}{8x(1-x)}+\frac{x(1-x)(\partial_x H^{(0)})^2}{2}\right]\right]\label{T1intintegral2}
\end{equation}
where the $j$-dependence has dropped out. At this stage the integrand contains terms (one inside $H^{(1)}$) that diverge at the edges and apparently prevent the $\epsilon\rightarrow0^+$ limit from being taken. However, after applying integration by parts to the $\theta$-integral of the second term these divergences are found to cancel. Taking the $\epsilon\rightarrow0^+$ limit then yields
\begin{equation}
	\fl \Ts^{(1)}_{I}=\int_0^1\dx\int_0^{2\pi}\frac{\dt}{2\pi}f'(H^{(0)})\left[\frac{1}{4}(2 x-1)(\partial_x H^{(0)}) +A_1(x)+2 B_1(x) \cos (\theta)\right].\label{T1intintegralfinal}
\end{equation}
\subsubsection{The edge terms $\Ts^{(1)}_{L}$ and $\Ts^{(1)}_{R}$}
The smoothness condition contained in \eref{ABdef} allows the edge corrections to be calculated using a simple combinatoric argument which is presented in the appendix. It is found that these corrections depend only on the values of $A_0(x)$ and $B_0(x)$ at $x=0,1$. In particular, $B_0(0)=0$ ($B_0(1)=0$) implies that $\Ts^{(1)}_{L}=0$ ($\Ts^{(1)}_{R}$=0) and the edge corrections therefore vanish if the sequence is closed. We define next
\begin{equation}
	\alpha(x)=A_0(x)-2B_0(x)\mt{and}{1}\beta(x)=A_0(x)+2B_0(x)
	\label{alphabetadef}
\end{equation}
and
\begin{equation}
	\tau_{[\alpha,\beta]}(\lambda)=\left\{\begin{array}{ll} \pi^{-1}\arcsin\left[\frac{\alpha+\beta-2\lambda}{\alpha-\beta}\right]  & \lambda\in[\alpha,\beta] \\ 0 & {\rm otherwise} \end{array} \right.
\end{equation}
If $B_0(0)>0$ the correction $\Ts^{(1)}_{L}$ can be expressed compactly as
\begin{equation}
	\Ts^{(1)}_{L}=\int_{-\infty}^{+\infty}{\rm d}x f(x) \frac{\rm d}{{\rm d}x}\tau_{[\alpha(0),\beta(0)]}(x)
	\label{edgecorrectionfinal}
\end{equation}
where it should be understood that the discontinuities of $\tau_{[\alpha(0),\beta(0)]}(x)$ at $x=\alpha(0),\beta(0)$ produce $\delta$-functions in the integrand. If $B_0(1)>0$ the correction $\Ts^{(1)}_{R}$ follows from \eref{edgecorrectionfinal} by replacing $[\alpha(0),\beta(0)]$ by $[\alpha(1),\beta(1)]$.\\

This concludes the derivation of the trace formula corresponding to the expansion of $\Ts_j$ in \eref{Tsexpand}. In the next section we show how detailed information regarding the asymptotic eigenvalue distribution of the matrix sequence follows from this result.
\section{The density of states}
\label{sectionDOS}
Let $\{\lambda(n,j)\}_{n=1}^{2j+1}$ denote the eigenvalues of $\hat{H}_j$. The scaled density of states associated with $\hat{H}_j$ is defined as
\begin{equation}
	\rho_j(\lambda)=\frac{1}{2j+1}\sum_{n=1}^{2j+1}\delta(\lambda-\lambda(n,j)).
	\label{dosdef}
\end{equation}
Our goal is to derive a linear order large-$j$ expansion of the form
\begin{equation}
	\rho_j(\lambda)=\rho^{(0)}(\lambda)+\rho^{(1)}(\lambda)/j+\oforder{j^{-2}}
\end{equation}
where $\rho^{(0)}(\lambda)$ and $\rho^{(1)}(\lambda)$ are defined by requiring that
\begin{eqnarray}
	\inflim{j}\int{\rm d}\lambda\,\rho_j(\lambda)g(\lambda)=\int{\rm d}\lambda\,\rho^{(0)}(\lambda)g(\lambda)\\
	\inflim{j}\int{\rm d}\lambda\,j[\rho_j(\lambda)-\rho^{(0)}(\lambda)]g(\lambda)=\int{\rm d}\lambda\,\rho^{(1)}(\lambda)g(\lambda)
\end{eqnarray}
holds for continuous $g(\lambda)$. It is useful to introduce the distribution functions associated with these asymptotic densities as 
\begin{equation}
	\Ds^{(0)}(\lambda)=\int_{-\infty}^\lambda{\rm d}\lambda'\,\rho^{(0)}(\lambda')\hs{1}{\rm and}\hs{1}\Ds^{(1)}(\lambda)=\int_{-\infty}^\lambda{\rm d}\lambda'\,\rho^{(1)}(\lambda').
	\label{Dasintegral}
\end{equation}
The traces considered in section \ref{sectiontraces} may now be expressed as
\begin{equation}
	\fl \Ts_j=\frac{1}{2j+1}{\rm tr}(f(\hat{H}_j))=\int{\rm d}\lambda\,\rho_j(\lambda)f(\lambda)\mt{and so}{0.8}\Ts^{(0,1)}=\int{\rm d}\lambda\,\rho^{(0,1)}(\lambda)f(\lambda).
	\label{traceassum}
\end{equation}
Knowledge of $\Ds^{(0,1)}(\lambda)$ is therefore sufficient to calculate $\Ts^{(0,1)}$ for any choice of $f(x)$. To determine $\Ds^{(0,1)}(\lambda)$ we first use this relation in reverse and express the density of states in terms of a particular trace, namely
\begin{equation}
	\Ts_j(\lambda,\gamma)=\frac{1}{2j+1}{\rm tr}\left[\Theta_\gamma(\lambda-\hat{H}_j)\right]
	\label{steptrace}
\end{equation}
where $\Theta_\gamma(x)=(1+\tanh(x/\gamma))/2$. Note that $\Theta_\gamma(x)$ is smooth for $\gamma>0$ but converges to the Heaviside step function $\Theta(x)$ as $\gamma\rightarrow0^+$. From \eref{Dasintegral} and \eref{traceassum} it follows that
\begin{equation}
	\Ds^{(0,1)}(\lambda)=\zeroplim{\gamma}\int{\rm d}\lambda'\,\rho^{(0,1)}(\lambda')\Theta_\gamma(\lambda-\lambda')=\zeroplim{\gamma}\Ts^{(0,1)}(\lambda,\gamma)
\end{equation}
and so from \eref{T0general} we find
\begin{equation}
	\Ds^{(0)}(\lambda)=\int_0^1\dx\int_0^{2\pi}\frac{\dt}{2\pi}\Theta(\lambda-H^{(0)}(x,\theta)).
	\label{D0integral2D}
\end{equation}
Performing the $\theta$-integral above yields the final form of $\Ds^{(0)}(\lambda)$ as
\begin{equation}
 \Ds^{(0)}(\lambda)=\int_0^1\dx\omega_{[\alpha(x),\beta(x)]}(\lambda)
 \label{D0integral}
\end{equation}
with $\alpha(x)$ and $\beta(x)$ as defined in \eref{alphabetadef} and 
\begin{equation}
	\omega_{[\alpha,\beta]}(\lambda)=\left\{\begin{array}{ll} 0 & \lambda<\alpha \\ 1 & \lambda>\beta \\ 2^{-1}+\pi^{-1}\arcsin\left[\frac{\alpha+\beta-2\lambda}{\alpha-\beta}\right]  &{\rm otherwise} \end{array} \right.
\end{equation}
From this it is evident that 
\begin{equation}
	\fl \lambda_-=\min\{\alpha(x)\,:\,x\in[0,1]\}\mt{and}{1}\lambda_+=\max\{\beta(x)\,:\,x\in[0,1]\}
	\label{lambdaPlusMin}
\end{equation}
defines the range of eigenvalues in the $j\rightarrow\infty$ limit. In particular, $\Ds^{(0)}(\lambda)=0$ for $\lambda\leq\lambda_-$ while $\Ds^{(0)}(\lambda)=1$ at $\lambda\geq\lambda_+$. Expression \eref{D0integral} was obtained by Kuijlaars and van Assche in \cite{kuijlaars1999}. In fact, this result is far more robust than the present derivation suggests; in \cite{kuijlaars2001} it was shown that \eref{D0integral} also holds for a wide class of discontinuous $\alpha(x)$ and $\beta(x)$. The expression for $\lambda_-$ as a ground state energy density was previously obtained in \cite{hollenberg1996}.\\

To treat the linear order correction $\Ds^{(1)}(\lambda)$ we first decompose it as
\begin{equation}
\fl \Ds^{(1)}(\lambda)=\zeroplim{\gamma}\left[\Ts^{(1)}_{L}(\lambda,\gamma)+\Ts^{(1)}_{I}(\lambda,\gamma)+\Ts^{(1)}_{R}(\lambda,\gamma)\right]=\Ds_{L}^{(1)}(\lambda)+\Ds_{I}^{(1)}(\lambda)+\Ds_{R}^{(1)}(\lambda).
	\label{D1sum}
\end{equation}
The interior contribution then follows from \eref{T1intintegralfinal} as
\begin{eqnarray}
\fl \Ds^{(1)}_{I}(\lambda)=\zeroplim{\gamma}\Ts^{(1)}_{I}&=&-\int_0^1\dx\int_0^{2\pi}\frac{\dt}{2\pi}\delta(\lambda-H^{(0)}(x,\theta))F(x,\cos(\theta))\label{D1bulkintegral2D}\\
	&=&-\int_0^1\dx\,\Omega_{[\alpha(x),\beta(x)]}(\lambda)\,F\left(x,(\lambda-A_0(x))/(2B_0(x))\right)
	\label{D1bulkintegral}
\end{eqnarray}
with $F(x,\cos(\theta))=(2 x-1)(\partial_x H^{(0)})/4 +A_1(x)+2 B_1(x)\cos(\theta)$. Here
\begin{equation}
	\Omega_{[\alpha,\beta]}(\lambda)=\left\{\begin{array}{ll} \pi^{-1}\left[(\lambda-\alpha)(\beta-\lambda)\right]^{-1/2} & \lambda\in(\alpha,\beta) \\ 0&{\rm otherwise} \end{array} \right.
	\label{Omegadefinition}
\end{equation}
when $\alpha<\beta$ while $\Omega_{[\alpha,\alpha]}(\lambda)=\delta(\lambda-\alpha)$. Note that $\Ds^{(1)}_{I}(\lambda)$ is zero outside the interval $[\lambda_-,\lambda_+]$. It only remains to determine the two edge corrections to $\Ds^{(1)}(\lambda)$ which are given by the $\gamma\rightarrow0^+$ limits of $\Ts^{(1)}_{L}(\lambda,\gamma)$ and $\Ts^{(1)}_{R}(\lambda,\gamma)$ for the trace in \eref{steptrace}. It follows from \eref{edgecorrectionfinal} that 
\begin{equation}
\fl 	\Ds^{(1)}_{L}(\lambda)=\frac{1}{4\pi}\arcsin\left[\frac{\lambda-A_0(0)}{2B_0(0)}\right]\hs{0.5}{\rm and}\hs{0.5}\Ds^{(1)}_{R}(\lambda)=\frac{1}{4\pi}\arcsin\left[\frac{\lambda-A_0(1)}{2B_0(1)}\right]
	\label{D1edge}
\end{equation}
for $\lambda\in[\alpha(0),\beta(0)]$ and $\lambda\in[\alpha(1),\beta(1)]$ respectively. Outside these ranges both corrections are zero. Also, if $B_0(0)=0$ ($B_0(1)=0$) then $\Ds^{(1)}_{L}(\lambda)$ ($\Ds^{(1)}_{R}(\lambda)$) is identically zero.\\

To summarize, equations \eref{D0integral}, \eref{D1sum}, \eref{D1bulkintegral} and \eref{D1edge} together yield expressions for $\Ds^{(0,1)}(\lambda)$. Taking derivatives to $\lambda$ produces $\rho^{(0,1)}(\lambda)$ and by using \eref{traceassum} the corrections $\Ts^{(0,1)}$ for an arbitrary trace can be calculated. In this regard, note that the linear order correction $\Ds^{(1)}(\lambda)$ is generally not continuous at $\lambda=\lambda_\pm$ and that this gives rise to $\delta$-functions in the density of states.
\section{Further results for special cases}
\label{sectionspecialcases}
In this section we consider matrix sequences for which certain quantities associated with the eigenstates are well-behaved functions of the corresponding eigenvalue or state label. The conditions under which these requirements are met will be investigated in the context of a specific example in section \ref{sectiondw}.
\subsection{The index function}
If the eigenvalues of $\hat{H}_j$ satisfy $\lambda(n,j)<\lambda(n+1,j)$ we define the index function as
\begin{equation}
	\Is(\lambda(n,j),j)=\frac{n}{2j+1},
	\label{Isdefinition}
\end{equation}
which is simply the fraction of states with energies less than \emph{or equal to} $\lambda(n,j)$. Note that $\lambda(n,j)$ and $\Is(\lambda(n,j),j)$ are, with respect to their first arguments, essentially inverses of each other. We again seek asymptotic expansions of these functions in the form
\begin{eqnarray}
	\Is(\lambda(n,j),j)&=&\Is^{(0)}(\lambda(n,j))+\Is^{(1)}(\lambda(n,j))/j+\oforder{j^{-2}} \label{Isasymptotic}\\
	\lambda(n,j)&=&\lambda^{(0)}(n/D)+\lambda^{(1)}(n/D)/j+\oforder{j^{-2}} \label{Dsasymptotic}
\end{eqnarray}
where $D=2j+1$. Let $\{n_j\}$ be a sequence of labels such that \mbox{$\inflim{j} n_j/(2j+1)=x\in[0,1]$}. We then define
\begin{equation}
\fl \lambda^{(0)}(x)=\inflim{j}\lambda(n_j,j)\mt{and}{1}\lambda^{(1)}(x)=\inflim{j}j\left[\lambda(n_j,j)-\lambda^{(0)}(n_j/(2j+1))\right].
\label{lam0lam1limits}
\end{equation}
Similarly, if $\{n_j\}$ is such that $\inflim{j}\lambda(n_j,j)=\lambda$ define
\begin{equation}
\fl \Is^{(0)}(\lambda)=\inflim{j}\frac{n_j}{2j+1}\mt{and}{1}\Is^{(1)}(\lambda)=\inflim{j}j\left[\frac{n_j}{2j+1}-\Is^{(0)}(\lambda(n_j,j))\right].
\label{I0I1limits}
\end{equation}
As to the existence and uniqueness of these limits we can say the following. First, it is clear that $\Is^{(0)}(\lambda)$ equals $\Ds^{(0)}(\lambda)$ and that $\lambda^{(0)}(x)$ is simply its inverse. A necessary condition for the existence of $\lambda^{(1)}(x)$ is then that
\begin{equation}
	\inflim{j} j\left[\lambda(n_j+m,j)-\lambda(n_j,j)\right]=m/(2\rho^{(0)}(\lambda))
	\label{gaplimit}
\end{equation}
must hold for any integer $m$. For $j$ sufficiently large the gap \mbox{$\Delta_{n,j}\equiv\lambda(n+1,j)-\lambda(n,j)$} between successive eigenvalues  must therefore behave as $\Delta_{n,j}\approx(2j\rho^{(0)}(\lambda(n,j)))^{-1}$ and be a slowly varying function of $n$ in the sense that $(\Delta_{n+1,j}-\Delta_{n,j})=\mathcal{O}(j^{-2})$. In section \ref{sectiondw} we consider a matrix sequence for which this condition is met only in certain regions of the spectrum. We will then also postulate a sufficient condition for the existence of $\lambda^{(1)}(x)$ and $\Is^{(1)}(x)$.\\

Assuming that the expansions in \eref{Isasymptotic} and \eref{Dsasymptotic} are valid we proceed to investigate the relation between $\Is^{(1)}(\lambda)$ and $\Ds^{(1)}(\lambda)$. First note that
\begin{equation}
		\Is(\lambda(n_j,j),j)=\frac{n_j}{2j+1}=\zeroplim{\gamma}\frac{{\rm tr}\left[\Theta_\gamma(\lambda(n_j,j)-\hat{H}_j)\right]+1/2}{2j+1}
\end{equation}
with $\Theta_\gamma(x)=(1+\tanh(x/\gamma))/2$ and where $1/2$ has been added to compensate for the fact that $\Theta_{\gamma}(0)=1/2$. Combining this representation of $\Is(\lambda(n_j,j),j)$ with \eref{Isasymptotic}, \eref{Dsasymptotic} and the results of section \ref{sectionDOS} leads to
\begin{equation}
		\Is^{(0)}(\lambda)=\Ds^{(0)}(\lambda)\hs{1}{\rm and}\hs{1}\Is^{(1)}(\lambda)=\Ds^{(1)}(\lambda)+\frac{1}{4}
		\label{IitoD}
\end{equation}
for $\lambda\in[\lambda_-,\lambda_+]$. This agrees with the expectation that the eigenvalue distribution function is almost exactly the fraction of states with energies at or below $\lambda$. The significance of the additional $1/4$ term will become clear when we compare these results to numerical data.  In cases where $\Is^{(1)}(\lambda)$ does not exist in the sense of \eref{I0I1limits} we may interpret $\Ds^{(1)}(\lambda)+1/4$ as representing $\Is^{(1)}(\lambda)$ in an average sense. 
\subsection{Expectation values and eigenstates}
\label{sectionwf}
In this section we work to lowest order in $1/j$. Consider a sequence of tridiagonal Hermitian matrices $\{\hat{Q}_j\}$ defined by the functions $\tilde{A}_0(x)$ and $\tilde{B}_0(x)$ as in \eref{ABdef}. Let $\ket{\lambda(n,j)}$ denote the eigenstate of $\hat{H}_j$ with eigenvalue $\lambda(n,j)$ and expansion coefficients \mbox{$\{c_m=\inp{j,-j+m}{\lambda(n,j)}\}_{m=0}^{2j}$} in the $\hat{J}_z$-basis. The expectation value $\matel{\lambda(n,j)}{\hat{Q}_j}{\lambda(n,j)}$ can then be expressed as
\begin{equation}
	\matel{\lambda(n,j)}{\hat{Q}_j}{\lambda(n,j)}=\int{\rm d}x\tilde{A}_0(x)\psi_{n,j}(x)+2\int{\rm d}x\tilde{B}_0(x)\phi_{n,j}(x)
\end{equation}
where 
\begin{equation}
	\fl \psi_{n,j}(x)=\sum_{m=0}^{2j}c_m^2\delta\left[x-\frac{m}{2j}\right]\hs{1}{\rm and}\hs{1}\phi_{n,j}(x)=\sum_{m=0}^{2j-1}c_{m+1}c_m\delta\left[x-\frac{2m+1}{4j}\right].
	\label{psiphifinitej}
\end{equation}
These functions encode information regarding the norm and relative signs of the eigenstate's expansion coefficients. Let $\{n_j\}$ be a sequence of state labels such that $\inflim{j}\lambda(n_j,j)=\lambda$. We will assume that the weak limits $\psi_{\lambda}(x)$ and $\phi_{\lambda}(x)$ of $\psi_{n_j,j}(x)$ and $\phi_{n_j,j}(x)$ exist and are determined by $\lambda$ alone. The same will then hold for the expectation value, and we may define 
\begin{equation}
	\langle\hat{Q}\rangle(\lambda)=\inflim{j}\matel{\lambda(n_j,j)}{\hat{Q}_j}{\lambda(n_j,j)}.
	\label{Qaslimit}
\end{equation}
The ``wave functions" $\psi_{\lambda}(x)$ and $\phi_{\lambda}(x)$ can be related to the asymptotic distribution of eigenvalues as follows. Consider the matrix sequence $\{\hat{H}_j+\epsilon\hat{Q}_j\}$ and denote the $n$'th eigenvalue of $\hat{H}_j+\epsilon\hat{Q}_j$ by $\lambda_{\epsilon}(n,j)$. The Hellmann-Feynman theorem now implies that
\begin{equation}
	\fl \matel{\lambda(n,j)}{\hat{Q}_j}{\lambda(n,j)}=\left.\frac{{\rm d}\lambda_\epsilon(n,j)}{{\rm d}\epsilon}\right|_{\epsilon=0}=\int{\rm d}x\tilde{A}_0(x)\frac{\delta\lambda(n,j)}{\delta A_0(x)}+\int{\rm d}x\tilde{B}_0(x)\frac{\delta\lambda(n,j)}{\delta B_0(x)}.
\end{equation}
Working to lowest order in $1/j$ we replace $\lambda(n,j)$ by $\lambda^{(0)}(n/D)$ and note that from  \eref{Isasymptotic}, \eref{Dsasymptotic} and \eref{IitoD} we have $n/D=\Is^{(0)}(\lambda^{(0)}(n/D))=\Ds^{(0)}(\lambda^{(0)}(n/D))$. Taking functional derivatives to $A_0(x)$ on both sides of the latter equation then leads to 
\begin{equation}
	\frac{\delta\lambda^{(0)}(n/D)}{\delta A_0(x)}=\left.\frac{-1}{\rho^{(0)}(\lambda)}\frac{\delta\Ds^{(0)}(\lambda)}{\delta A_0(x)}\right|_{\lambda=\lambda^{(0)}(n/D)}
\end{equation}
with a similar result for $B_0(x)$. Performing the functional derivatives produce
\begin{equation}
	\langle\hat{Q}\rangle(\lambda)=\int{\rm d}x\tilde{A}_0(x)\psi_\lambda(x)+2\int{\rm d}x\tilde{B}_0(x)\phi_\lambda(x)
	\label{expvalgeneral}
\end{equation}
where
\begin{equation}
	\fl \psi_\lambda(x)=\frac{1}{\rho^{(0)}(\lambda)}\Omega_{[\alpha(x),\beta(x)]}(\lambda)\hs{1}{\rm and}\hs{1}\phi_\lambda(x)=\frac{\lambda-A_0(x)}{2B_0(x)\rho^{(0)}(\lambda)}\Omega_{[\alpha(x),\beta(x)]}(\lambda).
	\label{wavefunctions}
\end{equation}
with $\Omega_{[\alpha,\beta]}(\lambda)$ as defined in \eref{Omegadefinition}. Note that $\psi_\lambda(x)$ and $\phi_\lambda(x)$ are supported on the set \mbox{$\Ss(\lambda)=\{x : \alpha(x)\leq\lambda\leq\beta(x)\}$}.
\subsection{Example: Alternating States}
\label{sectiondw}
\begin{figure}[t]
\begin{tabular}{cc}
\hs{-0.2}\imagetop{\includegraphics[width=7.6cm]{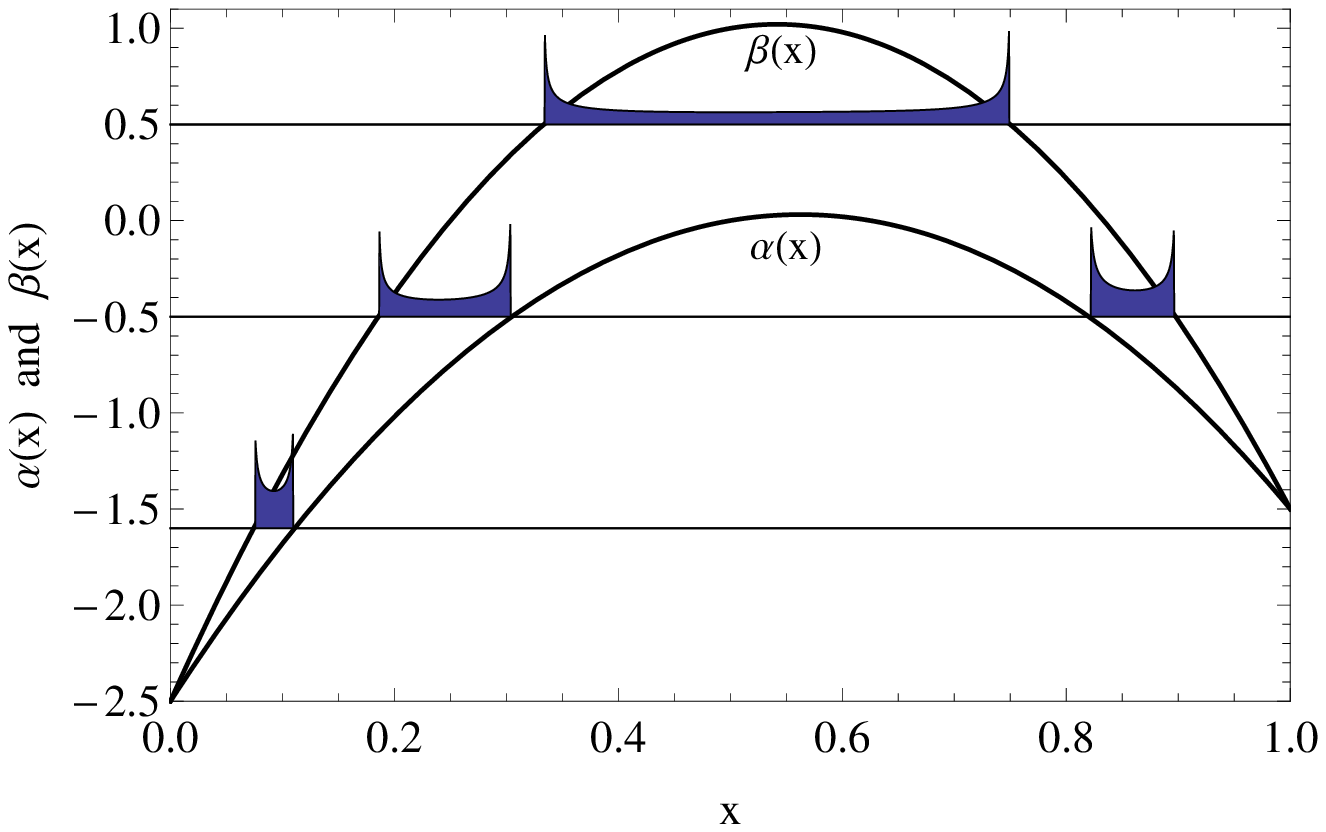}} & \imagetop{\includegraphics[width=7.6cm]{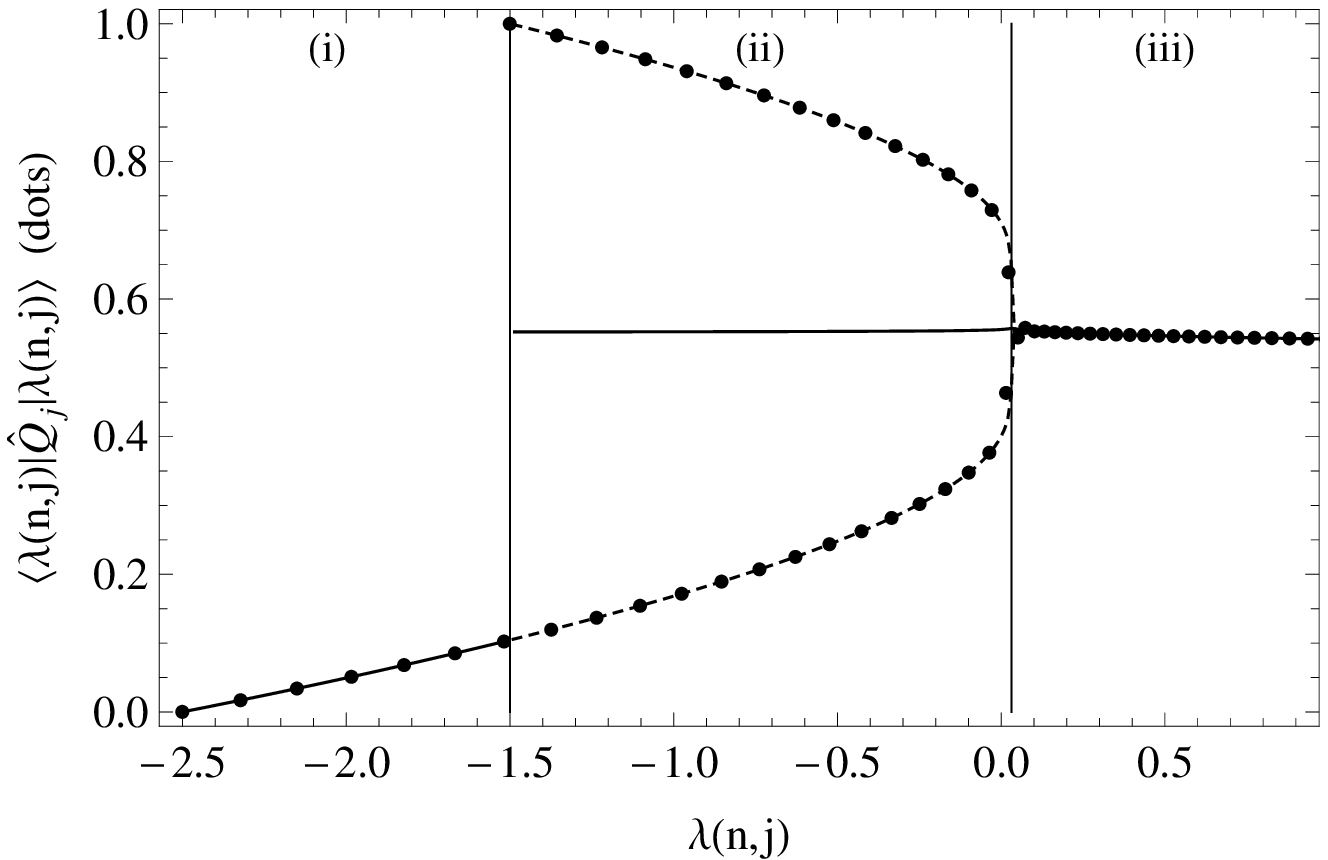}}\\
(a) & (b)\\
\end{tabular}
\caption{(a) $\alpha(x)$ and $\beta(x)$ for the example considered in section \ref{sectiondw}. The form of $\psi_\lambda(x)$ at $\lambda=-8/5, -1/2, 1/2$ is also shown. (b) Numerical expectation values for \mbox{$\hat{Q}_j=\hat{J}_z/(2j)+1/2$} with respect to the eigenstates $\ket{\lambda(n,j)}$ of $\hat{H}_j$. Here $j=30$. The solid line is the result of \eref{expvalgeneral}. The dashed lines are obtained from \eref{expvalgeneral} by restricting the integration domain to one of the intervals that define the support of $\psi_\lambda(x)$.}
\label{dwfigures}
\end{figure}
A general sufficient condition for $\Is^{(1)}(\lambda)$ and $\langle\hat{Q}\rangle(\lambda)$ to be well-defined in terms of the limits in \eref{I0I1limits} and \eref{Qaslimit} is still lacking. In this section we investigate this question in the context of a specific example. The insight gained from this special case will suggest a simple condition under which the results of the previous two sections may be applicable.\\

Consider the sequence defined by \mbox{$A_0(x)=-10x^2+11x-5/2$}, $B_0(x)=x(1-x)$, $A_1(x)=0$ and $B_1(x)=0$. It was found that the functions $\psi_\lambda(x)$ and $\phi_\lambda(x)$ are supported on $\Ss(\lambda)$ and encode information regarding the asymptotic structure of the eigenstates. The form of $\Ss(\lambda)$ therefore reveals how the states are localized in the $\hat{J}_z$ basis. Figure \ref{dwfigures} (a) shows $\alpha(x)$ and $\beta(x)$ as well as the shape of $\phi_\lambda(x)$ at three different energies. (See also figure \ref{lipkinfigures} (b) for a comparison with numerical results for the Lipkin model.) The spectrum can be divided into three regions based on the number of solutions to $\alpha(x)=\lambda$ and $\beta(x)=\lambda$: (i) $\lambda<-3/2$, (ii) $-3/2\leq\lambda<1/32$ and (iii) $1/32\leq\lambda\leq1$ where $1/32$ is the maximum value of $\alpha(x)$. In regions (i) and (iii) we see that $\Ss(\lambda)$ is a single interval while in (ii) it is the disjoint union of two intervals. Since this sequence is closed these intervals are always terminated by solutions to $\alpha(x)=\lambda$ and $\beta(x)=\lambda$.\\

An important fact which is not reflected by the asymptotic results is that at finite $j$ the eigenstates in region (ii) are localised alternatingly in the two intervals which form $\Ss(\lambda)$, but never in both simultaneously. This detail is lost in the large $j$-limit which effectively averages the properties of neighbouring eigenstates. This is also seen in the numeric results for the expectation value of $\hat{Q}_j=\hat{J}_z/(2j)+1/2$ shown in figure \ref{dwfigures} (b). In regions (i) and (iii) the expectation value varies slowly as a function of $\lambda(n,j)$ but alternate between two branches in region (ii). The prediction of \eref{expvalgeneral} and \eref{wavefunctions} is indicated by a solid line and, while agreeing with the numerical results in regions (i) and (iii), clearly produces an average value for $\matel{\lambda(n,j)}{\hat{Q}_j}{\lambda(n,j)}$ in region (ii). The correct expectation value along a single branch may be found by restricting the integral in \eref{expvalgeneral} to one of the two intervals that form $\Ss(\lambda)$. To preserve normalisation the same restriction is required in the integral for $\rho^{(0)}(\lambda)$. The results of this procedure appear as dashed lines in the figure. Similarly, we note that in region (ii) the gap $\Delta_{n,j}\equiv\lambda(n+1,j)-\lambda(n,j)$ also alternates between two branches and is therefore not a slowly varying function of $n$ as required by \eref{gaplimit}.\\

We are led to conclude that in region (ii) the quantities $\Is^{(1)}(\lambda)$ and $\langle \hat{Q}\rangle(\lambda)$ are not well-defined in terms of the limits in \eref{I0I1limits} and \eref{Qaslimit} since the results will depend on the choice of the index sequence $\{n_j\}$. In this region the analytic expressions for $\Is^{(1)}(\lambda)$ and $\langle\hat{Q}\rangle(\lambda)$ in \eref{IitoD} and \eref{expvalgeneral} must be interpreted as representing averages over neighbouring states in the spectrum. Based on these observations we postulate that $\Is^{(1)}(\lambda)$ and $\langle \hat{Q}\rangle(\lambda)$ are well-defined according to \eref{I0I1limits} and \eref{expvalgeneral} for regions of the spectrum where $\Ss(\lambda)$ is a single interval. This phenomenon was also noted in \cite{ribeiro2007,ribeiro2008} in the context of the Lipkin model.

\section{Band diagonal matrices}
\label{sectionbanddiagonal}
The integral expressions for $\Ds^{(0)}(\lambda)$ and $\Ds^{(1)}(\lambda)$ in \eref{D0integral2D} and \eref{D1bulkintegral2D} can easily be generalized to apply to closed sequences of Hermitian matrices with a fixed number $M$ of smooth off-diagonal bands. The matrix elements, from top left to bottom right, of the $m$'th lower band is denoted
\begin{equation}
	a_{n,j}^{(m)}=c^{(m)}_{n,j}+i\,d^{(m)}_{n,j}\mt{for}{1}n=m,\ldots,2j
\end{equation}
where $d_{n,j}^{(0)}=0$. The matrix structure is
\begin{equation}
\hat{H}_j=\left[
\begin{array}{ccccccccc}
a^{(0)}_{0} & \bar{a}^{(1)}_{1} & \bar{a}^{(2)}_{2} & \cdots & \bar{a}^{(M)}_{M} & 0     & \cdots & 0 \\
a^{(1)}_{1} & a^{(0)}_{1} & \bar{a}^{(1)}_{2} & \ddots & \ddots & \bar{a}^{(M)}_{M+1} & \ddots & \vdots \\
a^{(2)}_{2} & a^{(1)}_{2} & a^{(0)}_{2} & \ddots & \ddots & \ddots & \ddots & 0 \\
\vdots & \ddots & \ddots & \ddots & \ddots & \ddots & \ddots & \bar{a}^{(M)}_{2j} \\
a^{(M)}_{M} & \ddots & \ddots & \ddots & \ddots & \ddots & \ddots & \vdots \\
0     & a^{(M)}_{M+1} & \ddots & \ddots & \ddots & \ddots & \ddots & \bar{a}^{(2)}_{2j} \\
\vdots & \ddots & \ddots & \ddots & \ddots & \ddots & \ddots & \bar{a}^{(1)}_{2j} \\
0     & \cdots & 0     & a^{(M)}_{2j} & \cdots & a^{(2)}_{2j} & a^{(1)}_{2j} & a^{(0)}_{2j} \\
\end{array}
\right]
\end{equation}
where, for compactness, we have dropped the $j$ subscript and use $\bar{a}^{(m)}_{n}\equiv(a^{(m)}_{n})^*$. The smoothness condition in \eref{ABdef} now becomes
\begin{eqnarray}
	c^{(m)}_{n,j}=C_0^{(m)}\left(\frac{2n-m}{4j}\right)+\frac{1}{j}C_1^{(m)}\left(\frac{2n-m}{4j}\right)\\
	d^{(m)}_{n,j}=D_0^{(m)}\left(\frac{2n-m}{4j}\right)+\frac{1}{j}D_1^{(m)}\left(\frac{2n-m}{4j}\right)
	\label{generalsmooth}
\end{eqnarray}
and closure requires that $C_0^{(m)}(x)=D_0^{(m)}(x)=0$ for $x=0,1$ and $m>0$. Now define $\Fs_0(x,\theta)$ and $\Fs_1(x,\theta)$ by
\begin{equation}
	\Fs_{k}(x,\theta)=C^{(0)}_k(x)+2\sum_{m=1}^M\left[C_{k}^{(m)}(x)\cos(m\theta)+D_{k}^{(m)}(x)\sin(m\theta)\right]
	\label{Fsdefinition}
\end{equation}
Repeating the arguments of section \ref{sectionsmoothsymbols} leads us to conclude that $H^{(0)}(x,\theta)=\Fs_{0}(x,\theta)$ while
\begin{equation}
	H^{(1)}(x,\theta)=\Fs_{1}(x,\theta)+\frac{1}{16(1-x)x}\partial_\theta^2 \Fs_{0}(x,\theta)+\frac{(1-x)x}{4}\partial_x^2\Fs_{0}(x,\theta).
\end{equation}
Since the sequence is closed there are no edge corrections and the arguments of sections \ref{sectiontraces} and \ref{sectionDOS} can be employed virtually unchanged to arrive at general expressions for the lowest and linear order contributions to the asymptotic eigenvalue distribution:
\begin{eqnarray}
	\fl \Ds^{(0)}(\lambda)&=&\int_0^1\dx\int_0^{2\pi}\frac{\dt}{2\pi}\Theta(\lambda-\Fs_0(x,\theta))\\
\fl \Ds^{(1)}(\lambda)&=&\int_0^1\dx\int_0^{2\pi}\frac{\dt}{2\pi}\delta(\lambda-\Fs_0(x,\theta))\left[\Fs_1(x,\theta)+\frac{2x-1}{4}\partial_x\Fs_{0}(x,\theta)\right].
	\label{D0D1general}
\end{eqnarray}
Taking functional derivatives to $C^{(m)}_0(x)$ and $D^{(m)}_0(x)$ now produces integral expressions for expectation values analogous to \eref{expvalgeneral} and \eref{wavefunctions}.
\section{Application I: Collective Spin models}
\label{sectioncollective}
Consider a collection of $N$ spin-1/2 systems interacting via an infinitely long-ranged interaction in the presence of an external magnetic field. Since all the spin pairs interact identically the Hamiltonian can be expressed in terms of collective degrees freedom which are just the three components of the total spin $\hat{\vec{S}}$. The most general Hamiltonian of this type is \cite{vidal2006}
\begin{equation}
	\hat{H}=\mathbf{h}\cdot\hat{\mathbf{S}}+g_x\hat{S}_x^2+g_y\hat{S}^2_y.
	\label{generalcollective}
\end{equation}
This model remains of great interest, particularly because it allows for the study of non-trivial quantum critical phenomenon in a simple setting. Here we will demonstrate how such models can be treated within our formalism in a simple and straightforward manner. For discussions of the physical phenomenon see \cite{vidal2006,ribeiro2007,ribeiro2008,liberti2010} and references therein.\\

In the ferromagnetic phase where $g_x,g_y<0$ the ground state belongs to the subspace on which $\hat{\mathbf{S}}^2=s(s-1)$ with $s=N/2$. Focusing on this irrep of $su(2)$ the matrix representation of $\hat{H}$ in the $\hat{S}_z$-basis is seen to have at most two off-diagonal bands with matrix elements exhibiting the smoothness property of \eref{generalsmooth} with $j\equiv s$. Since the latter is inherited directly from the $su(2)$ generators it is clear that any Hamiltonian constructed in terms of these generators can be treated in exactly the same way. Furthermore, these sequences are always closed and the symbols can be calculated exactly, even at finite $j$, by using the representation of the generators as differential operators \cite{klauder1985}. After rescaling $g_{x,y}\rightarrow g_{x,y}/N$ and $\hat{H}\rightarrow\hat{H}/j$ we find according to \eref{Fsdefinition}
\begin{eqnarray}
	\Fs_0(x,\theta)&=&(2x-1)h_z+(1-x)x\left(g_x+g_y\right)-2\sqrt{(1-x)x} h_y \sin (\theta )\nonumber\\
	&&+2 \sqrt{(1-x)x}h_x\cos (\theta )+(1-x)x\left(g_x-g_y\right)\cos(2\theta)\\
	\Fs_1(x,\theta)&=&\frac{1}{4} \left[g_x+g_y+\left(g_x-g_y\right)\cos (2 \theta )+\frac{h_x \cos (\theta )-h_y \sin
   (\theta )}{\sqrt{(1-x)x}}\right].
\end{eqnarray}
Inserting these expressions into \eref{D0D1general} yields explicit integral expressions for the eigenvalue distribution function and its finite size corrections. Taking derivatives to the various coupling constants then produce expressions for expectation values, although in this regard the caveat discussed in section \ref{sectiondw} must be kept in mind. We will not embark on a study of the general Hamiltonian here, but instead highlight two prominent special cases. We return to the notation used for tridiagonal sequences prior to section \ref{sectionbanddiagonal}.

\subsection{Lipkin Model}
As a special case of the well-known LGM model \cite{lipkin1965}, itself a particular instance of \eref{generalcollective}, we consider
\begin{equation}
	\hat{H}=\hat{S}_z+\frac{\gamma}{4s}(\hat{S}_+^2+\hat{S}_-^2)
\end{equation}
within the $s=N/2$ sector. Note that $\hat{H}$ leaves the even \mbox{$\mathcal{H}_e={\rm span}\{\ket{s,-s+n}:n\ {\rm even}\}$} and odd \mbox{$\mathcal{H}_o={\rm span}\{\ket{s,-s+n}:n\ {\rm odd}\}$} subspaces invariant, and that its matrix representation in each of these is tridiagonal. We define $j=s/2$ and $j=(s-1)/2$ in the even and odd sector respectively. After rescaling the matrix elements by $2j$ to render them extensive we find that, in both sectors, $A_0(x)=2x-1$, $B_0(x)=\lambda(1-x)x$ and $A_1(x)=0$. The two sectors are therefore distinguished by $B_1(x)$ which is $B_1(x)=\lambda/8$ and $B_1(x)=\lambda(3-4x+4x^2)/8$ for the even and odd case respectively. Deriving the asymptotic distribution function $\Ds^{(0)}(\lambda)$ and its finite-size correction $\Ds^{(1)}(\lambda)$ is now simply a matter of evaluating the integrals in \eref{D0integral} and \eref{D1bulkintegral}. This can be done exactly with the final result given in terms of  elliptic integrals. The results agree with those obtained in \cite{ribeiro2007,ribeiro2008} based on an analysis of the zeros of the Majorana representation of the eigenstates.
We will not analyse these results further here, but only add the following comments.\\

Referring back to \eref{Omegadefinition} and \eref{wavefunctions} we see that since $(\lambda-\alpha(x))(\beta(x)-\lambda)$ is a fourth degree polynomial in $x$ the calculation of expectations values for observables defined by polynomial $\tilde{A}_0(x)$ and $\tilde{B}_0(x)$ using \eref{expvalgeneral} will involve evaluating elliptic integrals. These integrals are known to satisfy certain recurrence relations \cite{byrd1954} which aid in this calculation. For example, consider the observable $\hat{Q}_j=\hat{J}_z/(2j)+1/2$ for which $\tilde{A}_0(x)=x$ and $\tilde{B}_0(x)=0$. The moments of $\hat{Q}$ then satisfy
\begin{equation}
	\langle\hat{Q}^m\rangle(\lambda)=\frac{1}{2a_0(m-1)}\sum_{i=1}^4 a_i(2+i-2m)\,\langle\hat{Q}^{m-i}\rangle(\lambda)
	\label{lipkinrecursion}
\end{equation}
where $(a_0,a_1,a_2,a_3)=\left(4 \gamma ^2,-8 \gamma ^2,4 \left(\gamma ^2-1\right),4 (\lambda +1),-(\lambda +1)^2\right)$. This relation is exact in the thermodynamic limit and holds for all energies and values of the coupling constant. Once $\langle\hat{Q}\rangle(\lambda)$ and $\langle\hat{Q}^2\rangle(\lambda)$ are known from \eref{expvalgeneral} all the higher moments of $\hat{Q}$ therefore follow recursively from \eref{lipkinrecursion}.\\

\begin{figure}[t]
\begin{tabular}{cc}
\hs{-0.5}\includegraphics[width=7.8cm]{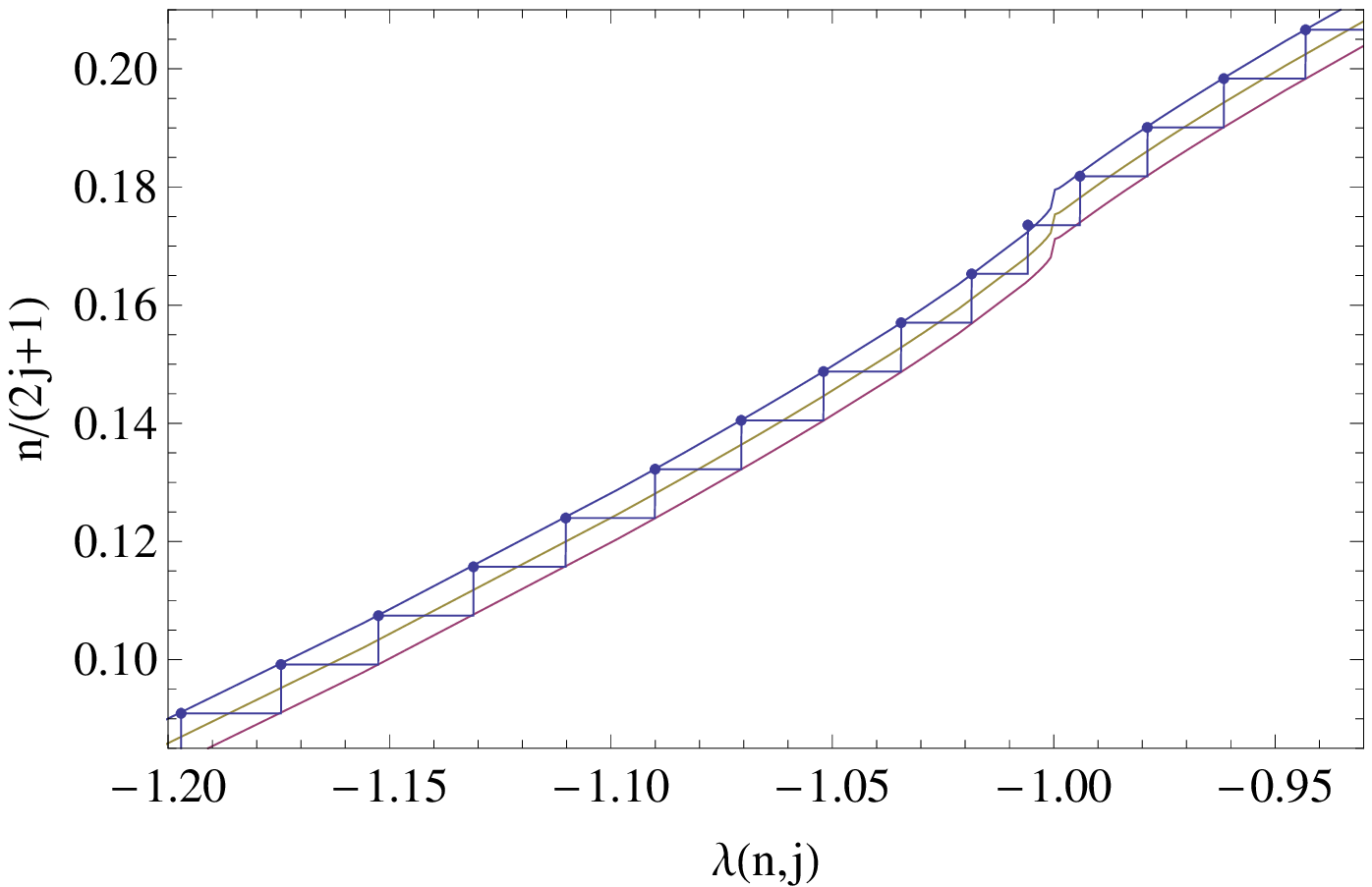} & \includegraphics[width=7.8cm]{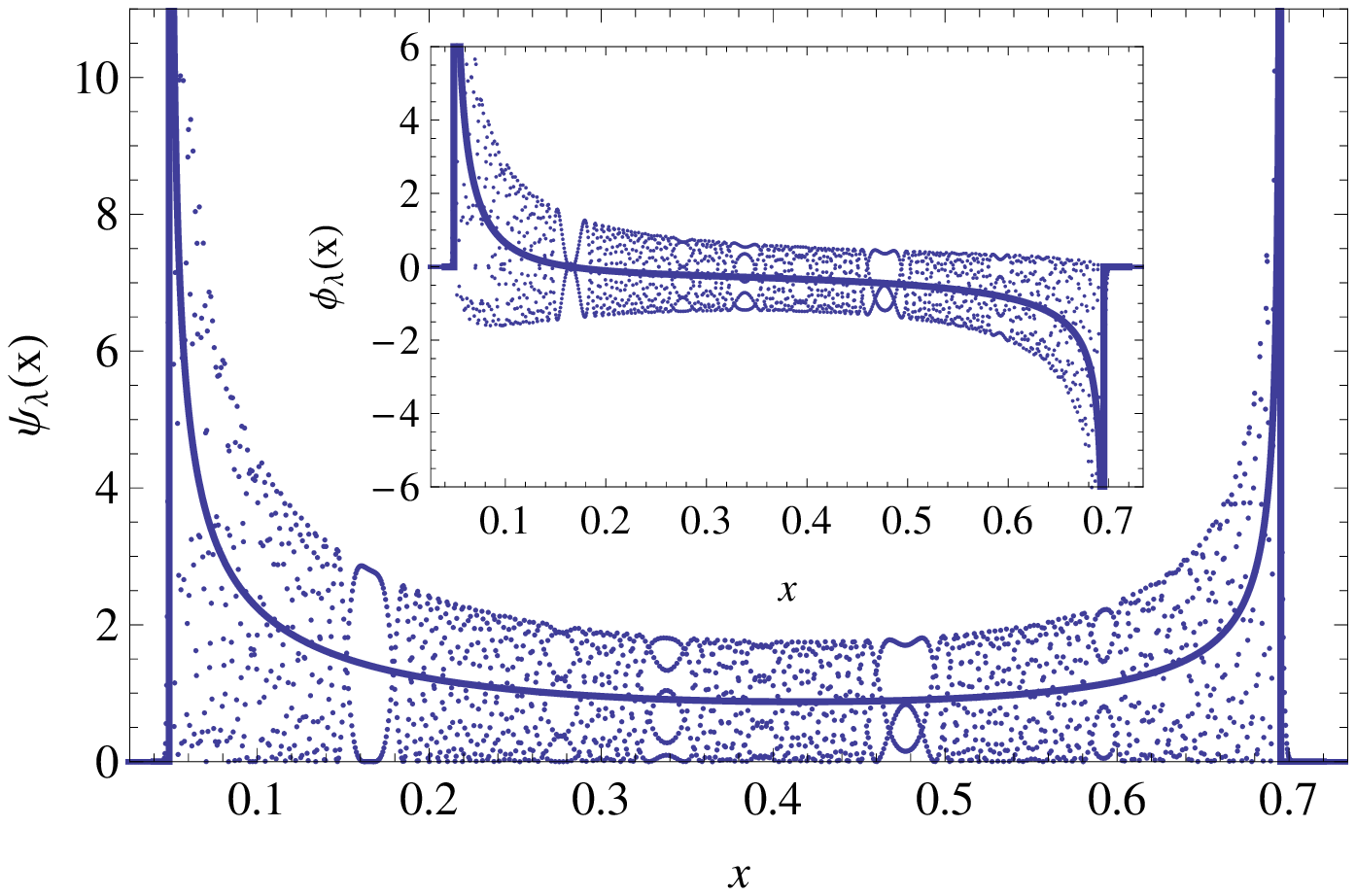}\\
(a) & (b)
\end{tabular}
\caption{(a) A comparison of analytic and numerical results for $j=60$ and $\gamma=2.5$. See text for details. (b) Plots of the functions $\psi_\lambda(x)$ and $\phi_\lambda(x)$ for the Lipkin model with $\gamma=2.5$ and $\lambda\approx-0.67$. Dots correspond to numerical data for $j=1500$. In the main plot the dots appear at $(m/(2j),2 j c_m^2)$ and in the inset at $((2m-1)/(4j),2 j c_m c_{m+1})$. The dots therefore represent the magnitudes of the $\delta$-functions in \eref{psiphifinitej}, but scaled by $2j$ to aid comparison with $\psi_\lambda(x)$ and $\phi_\lambda(x)$.
}
\label{lipkinfigures}
\end{figure}
Finally we perform some comparisons with numeric results. Let $\Is^{(L)}(\lambda)$ and $\Ds^{(L)}(\lambda)$ denote the linear order (in $1/j$) approximations to $\Is(\lambda)$ and $\Ds(\lambda)$. Keep in mind that from \eref{IitoD} it holds that  $\Is^{(L)}(\lambda)=\Ds^{(L)}(\lambda)+1/(4j)$. In figure \ref{lipkinfigures} (a) the three smooth curves, from top to bottom, are $\Is^{(L)}(\lambda)$, $\Ds^{(L)}(\lambda)$ and $\Ds^{(L)}(\lambda)-1/(4j)$. The dots indicate exact numerical values for $j=60$ while the staircase curve $S(\lambda)$ indicates the fraction of eigenvalues at or below a certain energy. The three smooth curves clearly represent slightly different large-$j$ approximations to $S(\lambda)$. In particular, we see why $\Is(\lambda)$, rather than $\Ds(\lambda)$, is the true reflection of the number of eigenvalues less than \emph{or equal to} a particular $\lambda(n,j)$. Figure \ref{lipkinfigures} (b) compares numerical results for $\psi_{n,j}(x)$ and $\phi_{n,j}(x)$ at $j=1500$ with $\psi_\lambda(x)$ and $\phi_\lambda(x)$.
\subsection{Uniaxial Model}
Another prominent special case of \eref{generalcollective} is the uniaxial model for which the Hamiltonian, in the symmetric phase \cite{liberti2010} and after appropriate rotations, reads
\begin{equation}
	\hat{H}=\hat{S}_x-\frac{4\gamma}{N}\hat{S}_z^2.
\end{equation}
Setting $j=s=N/2$ and rescaling the $\hat{H}$ by $j$ to render it intensive then produces the tridiagonal matrix sequence $\{\hat{H}_j\}$. The symbols can be calculated easily by representing the $su(2)$ generators as differential operators \cite{klauder1985}, and we find that \mbox{$A_0(x)=-2\gamma(1-2x)^2$}, $B_0(x)=\sqrt{(x-1)x}$, $A_1(x)=0$ and $B_1(x)=1/\sqrt{64(1-x)x}$. A myriad of information concerning the spectrum and expectation values, in both phases and for all energies, now follows from a straightforward application of the results derived in sections \ref{sectionDOS} and \ref{sectionspecialcases}. Here we only present new analytic results for the asymptotic density of states.\\

Using \eref{lambdaPlusMin} it is found that in the thermodynamic limit the eigenvalues of $\hat{H}_j$ range from
\begin{equation}
	\lambda_-(\gamma)=\left\{\begin{array}{cc} -1 &\ \ \gamma<1/4 \\ -2\gamma-1/(8\gamma) &\ \ {\rm \gamma\geq1/4} \end{array}\right.
\end{equation}
up to $\lambda_+=1$. We first consider the energy range $\lambda\in[-1,1]$. In terms of $A\equiv(1+8\gamma\lambda+16\gamma^2)^{1/2}$ and $B\equiv(A-1-4\gamma\lambda)/(2A)$ the density of states in the thermodynamic limit is $\rho^{(0)}(\lambda)=K(B)/(\sqrt{A}\pi)$ while its finite size correction reads
\begin{equation}
	\fl \rho^{(1)}(\lambda)=\frac{\delta(\lambda-1)}{4\sqrt{1+4\gamma}}+\Theta(1-4\gamma)\frac{\delta(\lambda+1)}{4\sqrt{1-4\gamma}}+\frac{\rm d}{{\rm d}\lambda}\left[\frac{2A E(B)-(1+A+8\lambda\gamma)K(B)}{8\pi\gamma\sqrt{A}}\right].
\end{equation}
Here $K(x)=F\left(\frac{\pi}{2}|x\right)$ and $E(x)=E\left(\frac{\pi}{2}|x\right)$ are complete elliptic integrals of the first and second kind \cite{byrd1954}.

Now consider $\lambda\in[\lambda_-(\gamma),-1]$ with $\gamma>1/4$. We set $C_\pm\equiv-1-4\lambda\gamma\pm4\gamma\sqrt{\lambda^2-1}$ and find that  $\rho^{(0)}(\lambda)=2K(C_-/C_+)/(\pi\sqrt{C_+})$ and
\begin{equation}
	\rho^{(1)}(\lambda)=\frac{2\gamma\,\delta(\lambda-\lambda_-)}{\sqrt{16\gamma^2-1}}+\frac{\rm d}{{\rm d}\lambda}\left[\frac{C_+ E(C_-/C_+)+(C_-+1)K(C_-/C_+)}{4\pi\gamma\sqrt{C_+}}\right].
\end{equation}
\section{Application II: Orthogonal Polynomials}
\label{sectionortpol}
It is well known that any sequence of orthogonal polynomials $\{p_m(x)\}_{m=0}^\infty$ satisfies a three-term recurrence relation \cite{gautschi2004} of the form 
\begin{equation}
	xp_m(x)=b_{m+1}p_{m+1}(x)+a_m p_{m}(x)+b_{m}p_{m-1}(x)
\end{equation}
where $b_m>0$. The symmetric, tridiagonal Jacobi matrix\footnote{Not to be confused with the $su(2)$ generators.} $\hat{J}_m$ is then defined in terms of the recurrence coefficients as $[\hat{J}_m]_{n,n}=a_{n-1}$ and $[\hat{J}_m]_{n+1,n}=[\hat{J}_m]_{n,n+1}=b_n$ for \mbox{$n=1,\ldots,m$.} Its eigenvalues are precisely the $m$ zeroes of $p_m(x)$. This simple correspondence enables the calculation of the asymptotic zero distributions plus corrections for a range of orthogonal polynomial sequences. Here we present only two well-known cases. Examples of other polynomial sequences to which this formalism can be applied may be found in \cite{kuijlaars1999,gautschi2004}.

\subsection{Laguerre Polynomials}
The recurrence coefficients for the generalized Laguerre polynomial $L_m^\alpha(x)$ are \mbox{$a_{n}=2n+\alpha+1$} and \mbox{$b_{n}=\sqrt{n(n+\alpha)}$} where $\alpha>-1$ is a real parameter. We define the matrix sequence $\{\hat{H}_j\}$ by $\hat{H}_j=\hat{J}_{2j+1}/j$. The $n$'th eigenvalue $\lambda(n,j)$ of $\hat{H}_j$ and root $\nu(n,j)$ of $L_{2j+1}^\alpha(x)$ are therefore related by $\lambda(n,j)=\nu(n,j)/j$. We also allow $\alpha$ to depend on $j$ as $\alpha=\alpha_0 j+\alpha_1$. The eigenvalues of $\hat{H}_j$ and the zeros of $L_{2j+1}^{(\alpha_0j+\alpha_1)}(xj)$ therefore share the same asymptotic distribution. The matrix elements of $\hat{H}_j$ are
\begin{equation}
 \fl a_{n,j}=(2n+\alpha_0 j+\alpha_1+1)/j\hs{1}{\rm and}\hs{1}b_{n,j}=\sqrt{n(n+\alpha_0 j+\alpha_1)}/j
\end{equation}
from which it follows by \eref{ABdef} that
\begin{eqnarray}
A_0(x)&=&4x+\alpha_0\hs{1}  B_0(x)=\sqrt{2x \left(2 x+\alpha_0\right)}\\
A_1(x)&=&1+\alpha_1\hs{1}  B_1(x)=\frac{4 x \left(\alpha_1+1\right)+\alpha_0}{4\sqrt{2x \left(2 x+\alpha_0\right)}}
\end{eqnarray}
Note that this sequence is not closed on the right since $B_0(1)>0$.\\

\begin{figure}[t]
\begin{tabular}{cc}
\hs{-0.2}\imagetop{\includegraphics[width=7.6cm,bb=0 0 400 235]{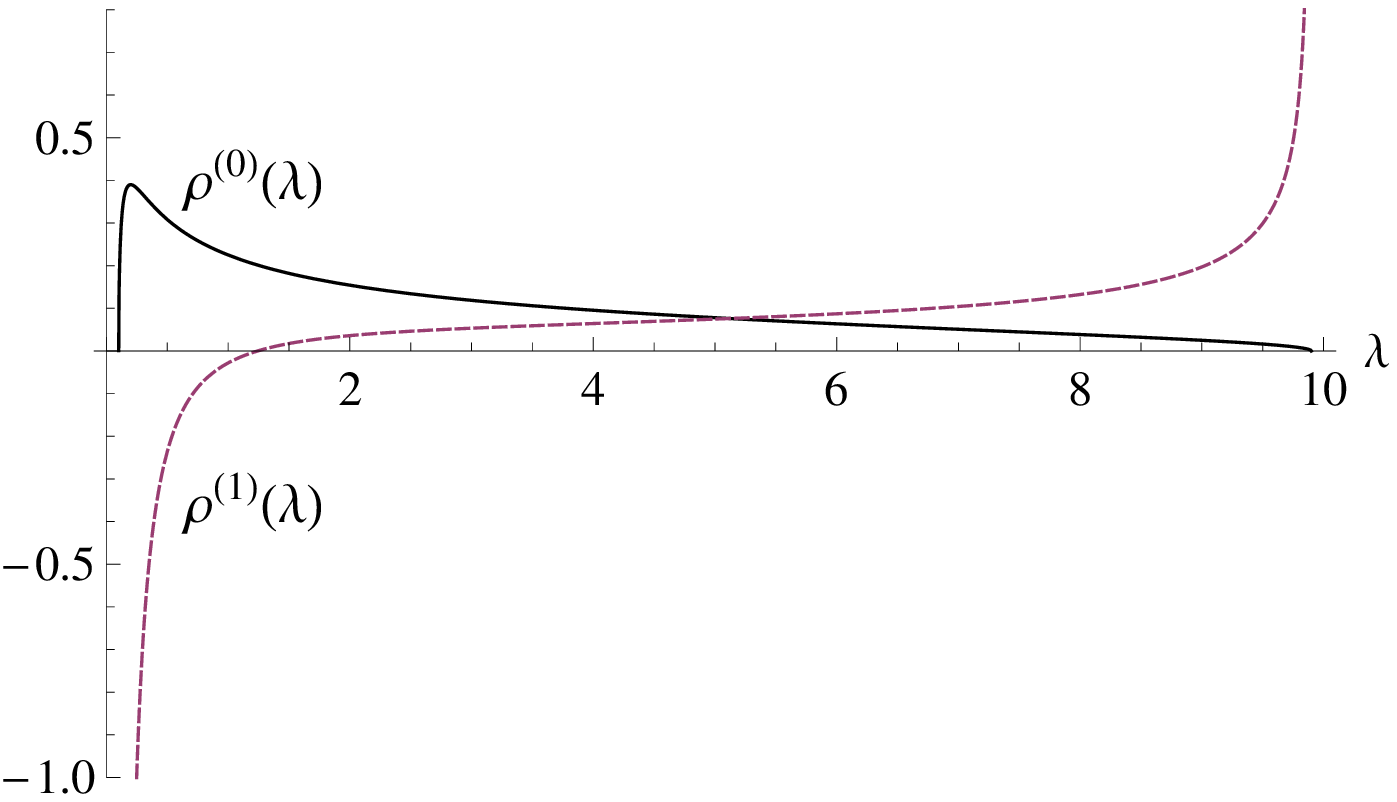}} & \imagetop{\includegraphics[width=7.6cm]{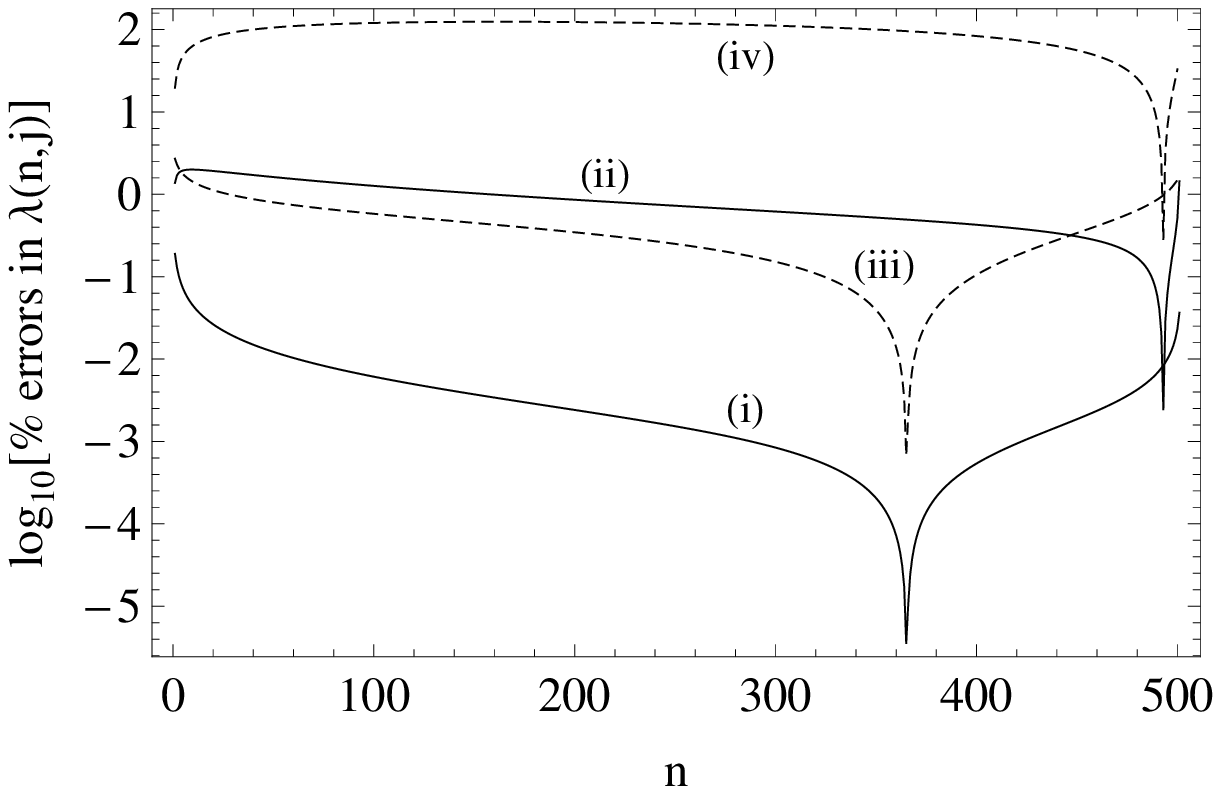}}\\
(a) & (b)
\end{tabular}
\caption{Results for the Laguerre polynomials with $\alpha_0=1$ and $\alpha_1=5$. (a) The lowest and linear order contributions to the asymptotic density of zeros. (b) Errors in the results of approximating $\lambda(n,j)$ by solving \eref{solveforlambda} for a given $n$ when $j=250$. Interpolating curves rather than individual points are shown. Solid curves (i) and (ii) show the percentage error when using the linear and lowest order approximations to $\Is_j(\lambda)$ respectively. Curve (iii) shows the error at linear order expressed as a percentage of $\lambda(n+1,j)-\lambda(n,j)$. Curve (iv) is as (iii) but for the lowest order approximation.}
\label{laguerrefigures}
\end{figure}

From this we can derive analytic expressions for $\Ds^{(0)}(\lambda)$ and $\Ds^{(1)}(\lambda)$ which appear in the appendix. The corresponding corrections to the density of zeros are
\begin{eqnarray}
	\rho^{(0)}(\lambda)&=&\frac{\sqrt{(\lambda_+-\lambda)(\lambda-\lambda_-)}}{4\pi\lambda}\\
	\rho^{(1)}(\lambda)&=&\frac{\left(\lambda -\alpha _0\right) \left(\lambda-\alpha _0+2\alpha _1-2\right)-2 \alpha _0}{8 \pi  \lambda  \sqrt{(\lambda_+-\lambda)(\lambda-\lambda_-)}}-\frac{1}{8}\delta(\lambda-\lambda_+)\nonumber\\
	&&-\frac{1}{8}\delta(\lambda-\lambda_-)\left(1+2\alpha_1\delta_{\alpha_0,0}\right)
\end{eqnarray}
which are supported on $\lambda\in[\lambda_-,\lambda_+]$ with $\lambda_\pm=4+\alpha_0\pm2\sqrt{4+2\alpha_0}$. Here $\delta_{a,b}$ is the Kronecker-delta. These results agree with those found in \cite{gawronski1993} for a closely related sequence. These functions are depicted in figure \ref{laguerrefigures} (a) for a specific choice of parameters. As a concrete measure of the accuracy of the asymptotic expansion we consider the problem of finding the $n$'th root of $L_{2j+1}^{(\alpha_0j+\alpha_1)}(x)$ for a given large $j$. This involves solving for $\lambda(n,j)$ numerically from
\begin{equation}
	\Is^{(0)}(\lambda(n,j))+\frac{1}{j}\Is^{(1)}(\lambda(n,j))=\frac{n}{2j+1}
	\label{solveforlambda}
\end{equation}
where $\Is^{(0,1)}(\lambda)$ is related to $\Ds^{(0,1)}(\lambda)$ by \eref{IitoD}. The errors in the resulting approximations to $\lambda(n,j)$ appear in figure \ref{laguerrefigures} (b) as a function of $n$ for $j=250$, $\alpha_0=1$ and $\alpha_1=5$. From curve (i) we see very good agreement with the exact results with errors being at worst $\approx 0.2\%$ and on average only $\approx 0.005\%$. Using the lowest order approximation (i.e. solving $\Is^{(0)}(\lambda(n,j))=n/(2j+1)$) produces errors of about two orders of magnitude greater (curve (ii)); in line with the fact that $j=250$. An insighful alternative measure of the accuracy is to compare the error in the approximation of $\lambda(n,j)$ with the difference between $\lambda(n,j)$ and $\lambda(n+1,j)$. This gives a indication of the ``resolving power" of this method, i.e. whether the approximation of a single root is sufficiently accurate to reliably distinguish it from its neigbours. When working up to lowest order this is not possible, as indicated by curve (iv); the error is of the same magnitude as the distance between succesive roots. However, to linear order (curve (iii)) we find that the errors are at most $\approx2.7\%$ and on average only $\approx0.25\%$ of this distance. Since the expressions for $\Is^{(0,1)}(\lambda)$ involve only elementary functions this procedure amounts to a simple and efficient numerical algorithm for accurately approximating arbitrary roots of high degree Laguerre polynomials. 

\subsection{Jacobi Polynomials}
The recurrence coefficients for the Jacobi polynomials $J_n^{(\alpha,\beta)}(x)$ are
\begin{eqnarray}
a_{n}(\alpha,\beta)&=&\frac{\beta^2-\alpha^2}{(2n+\alpha+\beta)(2+2n+\alpha+\beta)}\\
b_{n}(\alpha,\beta)&=&2\sqrt{\frac{n(n+\alpha)(n+\beta)(n+\alpha+\beta)}{(\alpha+\beta+2n-1)(2n+\alpha+\beta)^2(\alpha+\beta+2n+1)}}
\end{eqnarray}
where $\alpha>-1$ and $\beta>-1$ are real parameters. We again allow these parameters to vary with $j$ as $\alpha=\alpha_0 j+\alpha_1$ and $\beta=\beta_0 j+\beta_1$. The precise polynomial sequence is therefore $J_{2j+1}^{(\alpha_0 j+\alpha_1,\beta_0 j+\beta_1)}(x)$ while the corresponding matrix sequence $\{\hat{H}_j=\hat{J}_{2j+1}\}$ has elements
\begin{equation}
	\fl a_{n,j}=a_{n}(\alpha_0 j+\alpha_1,\beta_0 j+\beta_1) \hs{1}{\rm and}\hs{1}b_{n,j}=b_{n}(\alpha_0 j+\alpha_1,\beta_0 j+\beta_1).
	\label{jacobiME}
\end{equation}
Expressions for $\{A_{0,1},B_{0,1}\}$ follow easily from \eref{ABdef} and \eref{jacobiME}. Performing the integrals yields
\begin{eqnarray}
\rho^{(0)}(\lambda)&=&\frac{(\az+\bz+4)\sqrt{(\lambda_+-\lambda)(\lambda-\lambda_-)}}{4\pi(1-\lambda^2)}\\
\rho^{(1)}(\lambda)&=&\frac{\alpha_0+\beta_0+4}{4\pi\sqrt{(\lambda_+-\lambda)(\lambda-\lambda_-)}}\left[\frac{\alpha _0 \bar{\alpha}}{1-\lambda}+\frac{\beta _0 \bar{\beta}}{1+\lambda}-\frac{\left(\bar{\alpha}+\bar{\beta}-2\right)}{2(\alpha_0+\beta_0+4)}\right]\nonumber\\
   &&-\frac{1}{8}\delta(\lambda-\lambda_+)\left[1+2 \beta_1\delta_{\beta_0,0}\right]-\frac{1}{8}\delta(\lambda-\lambda_-)\left[1+2 \alpha_1\delta_{\alpha_0,0}\right]
\label{jacobirho}
\end{eqnarray}
where $\bar{\alpha}\equiv\alpha_0-2\alpha_1$ and $\bar{\beta}\equiv\beta_0-2\beta_1$. These densities are supported on $\lambda\in[\lambda_-,\lambda_+]$ with \mbox{$\lambda_\pm=A_0(1)\pm 2B_0(1)$}. These expressions differ from those obtained \cite{gawronski1991} for the same polynomial sequence. We have performed numerous numerical tests to verify our results and have not found any discrepancies. As in the case of the Laguerre polynomials analytic expressions for $\Is^{(0,1)}(\lambda)$ can be also be found. These allow for the implementation of the same numerical algorithm for finding specific roots of high degree Jacobi polynomials. Comparison with exact results show the same levels of accuracy as observed in the Laguerre case.\\

\section{Conclusion}
We have presented an approach to studying the asymptotic properties of a class of matrix sequences which arise in a number of diverse settings. Analytic expressions for the asymptotic eigenvalue distribution, eigenstates and expectation values were derived. Underpinning this approach is idea of reformulating matrix (or operator) problems in terms of their associated symbols for which the multiplication rule is given by the star product. A systematic expansion of the star product in orders of the inverse matrix dimension ($1/D$) then allowed for the derivation of a general trace formula which captures both the limiting $D\rightarrow\infty$  behaviour as well as linear order corrections. Once a particular matrix sequence has been specified through the functions $\{A_{0,1}(x),B_{0,1}(x)\}$ applying these results simply amounts to evaluating certain integral expressions. We considered applications to quantum spin systems and orthogonal polynomials and found that in both cases new and existing results can be obtained in a straightforward manner. In fact, in these two fields alone a range of applications are still to be investigated. A thorough investigation into applications in other fields should also prove fruitful.\\

Two aspects of our approach can benefit from further generalization. First, one would like to relax the closure restriction on the band diagonal case considered in section \ref{sectionbanddiagonal}. This will require a more general analysis of the edge corrections. Existing results regarding Toeplitz matrices may be of use in this regard. Secondly, one may consider the inclusion of higher order corrections going beyond the linear case. The existence of the star product expansion to all order, as reflected in expression \eref{starexpand}, suggests that this is certainly a possibility.

\section{Acknowledgements}
This work was supported under a grant of the National Research Foundation of South Africa. 

\appendix
\section{Edge corrections}
\label{appendixedge}
Here we derive expressions for the edge terms $\Ts^{(1)}_{L}$ and $\Ts^{(1)}_{R}$ appearing in the linear order correction to the trace. Focussing on $\Ts^{(1)}_{L}$ we see from \eref{T0finalline} and \eref{T1Integral} that
\begin{eqnarray}
	\Ts^{(1)}_{L}&=&\zeroplim{\epsilon}\inflim{j}\int_0^{\epsilon}\hs{0}\dx\int_0^{2\pi}\frac{\dt}{2\pi} j\left[\matel{z}{f(\hat{H}_j)}{z}-f(H^{(0)})\right].
\end{eqnarray}
Note that the $\theta$-integral acts as a projection which eliminates off-diagonal contributions in both $\matel{z}{f(\hat{H}_j)}{z}$ and $f(H^{(0)})$. Only the diagonal entries of $f(\hat{H}_j)$ are relevant, and similarly only the first term in the expansion $f(H^{(0)})=\sum_{n=0} g_n(x)\cos(n\theta)$ contributes. With this in mind we define $\hat{D}_j$ as the diagonal part of $f(\hat{H}_j)$ and $D^{(0)}(x)=(2\pi)^{-1}\int_0^{2\pi}\dt f(H^{(0)})$ 
in terms of which $\Ts^{(1)}_{L}$ reads
\begin{eqnarray}
	\Ts^{(1)}_{L}&=&\zeroplim{\epsilon}\inflim{j}\int_0^{\epsilon}\hs{0}\dx j\left[\matel{z}{\hat{D}_j}{z}-D^{(0)}(x)\right]
	\label{expr2}
\end{eqnarray}
with $z=\sqrt{x/(1-x)}$.\\

We first demonstrate the calculation of $\Ts^{(1)}_{L}$ for the case where $A_0(0)=0$ and \mbox{$f(x)=x^4$}. Let $d_0,\ldots,d_{2j}$ denote the diagonal matrix elements of $\hat{D}_j={\rm diag}(\hat{H}^4_j)$. Up to lowest order in $1/j$ it holds that
\begin{eqnarray}
	d_0&=&a_0^4+6a_0^2 b_1^2+2b_1^4+\oforder{1/j}\\
	d_1&=&a_1^4+12a_1^2 b_2^2+5b_2^4+\oforder{1/j}\\
	d_n&=&a_n^4+12a_n^2 b_{n+1}^2+6 b_{n+1}^4+\oforder{1/j}\hs{1}2\leq n\leq 2j-2
\end{eqnarray}
where the smoothness condition in \eref{ABdef} has been used to simplify the expressions by replacing, for example, $a_0+a_1$ by $2a_0+\oforder{1/j}$ and so on. Due to the tridiagonal structure of $\hat{H}_j$ the numeric coefficient of the $b^4_{n+1}$ term in $d_n$ has a simple combinatoric interpretation: it is the number of four step random walks on the lattice $\{1,2,3,\ldots\}$ which start and end at site $n+1$. Since only the walks associated with $d_0$ and $d_1$ are affected by the lattice boundary the numeric coefficients appearing in $d_0$ and $d_1$ do not conform to the general pattern observed in both $d_n$ for $2\leq n\leq 2j-2$ and in \mbox{$D^{(0)}(x)=A_0(x)^4+12 A_0(x)^2 B_0(x)^2+6 B_0(x)^4$}. 
(The coefficient of the $a_0^2 b_{1}^2$ term is similarly affected, although this will be seen not to influence $\Ts^{(1)}_{L}$.) It is this deviation of the numeric coefficients in $d_0$ and $d_1$ from those in $D^{(0)}(x)$ that gives rise to non-zero edge corrections. To make this explicit, let $\hat{D}'_j$ denote the diagonal matrix sequence in which these deviations in $d_0$ and $d_1$ have been ``corrected" to lowest order by defining $d'_n=d_n$ for $2\leq n\leq 2j-2$ and $d_0'=d_0+4b_1^4$ and $d_1'=d_0+b_2^4$. Note that no corrections are necessary for terms containing $a_n$ since $A_0(0)=0$ implies that $a_n=\oforder{1/j}$ if $n=\oforder{j^0}$. The matrix elements of $D'_j$ now represent, to lowest order in $1/j$, a sampling of the function $D^{(0)}(x)$ at the points $n/(2j)$ with $n=0,\ldots,2j$. It is clear that both $\{D'_j\}$ and $\{D_j\}$ converge to $D^{(0)}(x)$ on $x\in(0,1)$ but it also holds that $j[\matel{z}{\hat{D}'_j}{z}-D^{(0)}(x)]$ remains bounded as $j\rightarrow\infty$. Replacing $\hat{D}_j$ by $\hat{D}'_j$ in \eref{expr2} therefore allows the $j\rightarrow\infty$ limit to be taken into the integral after which taking $\epsilon\rightarrow0^+$ produces a zero result. It follows that
\begin{equation}
	\Ts^{(1)}_{L}=\zeroplim{\epsilon}\inflim{j}\int_0^{\epsilon}\hs{0}\dx j\matel{z}{\hat{D}_j-\hat{D}'_j}{z}.
	\label{expr2p5}
\end{equation}
Furthermore, since only the first two diagonal matrix elements of $\hat{D}_j-\hat{D}'_j$ are nonzero it holds that $j\matel{z}{\hat{D}_j-\hat{D}'_j}{z}$ will become completely localised around $x=0$ as $j\rightarrow\infty$ and so $\inflim{j}\int_0^{\epsilon}\hs{0}\dx j\matel{z}{\hat{D}_j-\hat{D}'_j}{z}$ is independent of $\epsilon\in(0,1)$. Combining these observations with the expression for the trace in \eref{traceasintegral} leads to 
\begin{equation}
	\Ts^{(1)}_{L}=\zeroplim{\epsilon}\inflim{j}\int_0^{\epsilon}\hs{0}\dx j\matel{z}{\hat{D}_j-\hat{D}'_j}{z}=\inflim{j}\frac{1}{2}\sum_{n=0}^{\lfloor \eta j\rfloor}(d_{n,j}-d'_{n,j})
	\label{expr3}
\end{equation}
where $\eta\in(0,1)$ is an arbitrary constant. The right hand side is readily evaluated to produce the left edge correction  
\begin{equation}
	\Ts^{(1)}_{L}=\inflim{j}(-4b_1^4-b_2^4)/2=-5(B_0(0))^4/2.
\end{equation}
This result can be generalised to apply to any $f(x)=x^m$ with $m$ even. Arguing as before the matrix elements of $\hat{D}'_j$ now become
\begin{eqnarray}
	d'_n&=&d_n+F_{m,n}b_{n+1}^m\hs{1} 0\leq n<m/2\\
	d'_n&=&d_n\hs{1} m/2\leq n
\end{eqnarray}
where $F_{m,n}$ is the number of $m$-step random walks on $\mathbb{Z}$ which start and end at $n+1$ and visit site zero at least once. A standard argument gives $F_{m,n}=\binom{m}{n+m/2}$ and expression \eref{expr3} produces the left edge correction for $f(x)=x^m$  as
\begin{equation}
	\Ts^{(1)}_{L}=-\frac{(B_0(0))^m}{2}\sum_{n=0}^{m/2}F_{m,n}=\frac{(B_0(0))^m}{4}\left[\binom{m}{m/2}-2^{m}\right]
	\label{Tseven}
\end{equation}
whenever $m$ is even. If $m$ is odd any correction terms added to $d_n$ to match the pattern in $D^{(0)}(x)$ will contain at least one $a_i$ factor and are therefore zero in the $j\rightarrow\infty$ limit when $A_0(0)=0$. $\Ts^{(1)}_{L}$ therefore vanishes whenever $f(x)$ is a odd power. For a general analytic $f(x)$ expressed as a power series the edge correction $\Ts^{(1)}_{L}$ will be a linear combination of the expressions in \eref{Tseven}. It would be preferable to have an expression for $\Ts^{(1)}_{L}$ directly in terms of $f(x)$ itself. For this purpose we note that
\begin{equation}
	\fl \int_{-1}^{+1}{\rm d}x\ (2x)^m \left[\frac{1}{\pi\sqrt{1-x^2}}-\frac{\delta(x-1)}{2}-\frac{\delta(x+1)}{2}\right]=\left\{\begin{array}{cc}0 &\  m\ {\rm odd}\\\binom{m}{m/2}-2^{m} &\  m\ {\rm even} \end{array}\right.
\end{equation}
and comparison with \eref{Tseven} then reveals that, for a general analytic $f(x)$, 
\begin{equation}
	\Ts^{(1)}_{L}=-f(-2B_0(0))/8-f(2B_0(0))/8+\int_{-1}^{+1}{\rm d}x\ \frac{f(2B_0(0)x)}{4\pi\sqrt{1-x^2}}.
	\label{Ts103}
\end{equation}
The only remaining restriction on this result is that $A_0(0)$ must be zero, but this can be circumvented as follows. If a given $\{\hat{H}_j\}$ does not satisfy $A_0(0)=0$ then the sequence $\{\hat{H}_j-A_0(0)\}$ surely does. Instead of applying \eref{Ts103} to the sequence $\{\hat{H}_j\}$ and function $f(x)$ we simply consider $\{\hat{H}_j-A_0(0)\}$ and $g(x)\equiv f(x+A_0(0))$ instead. Recalling that \mbox{$\alpha(x)=A_0(x)-2B_0(x)$} and $\beta(x)=A_0(x)+2B_0(x)$ the final expression for $\Ts^{(1)}_{L}$ can be written as
\begin{eqnarray}
	\fl\Ts^{(1)}_{L}&=&-f(\alpha(0))/8-f(\beta(0))/8+\int_{\alpha(0)}^{\beta(0)}{\rm d}x\ \frac{f(x)}{4\pi\sqrt{(\beta(0)-x)(x-\alpha(0))}}\\
	\fl&=&\int_{-\infty}^{+\infty}{\rm d}x f(x) \frac{\rm d}{{\rm d}x}\tau_{[\alpha(0),\beta(0)]}(x)
	\label{edgecorrectionfinalapp}
\end{eqnarray}
where
\begin{equation}
	\tau_{[\alpha,\beta]}(\lambda)=\left\{\begin{array}{ll} (4\pi)^{-1}\arcsin\left[\frac{\alpha+\beta-2\lambda}{\alpha-\beta}\right]  & \lambda\in[\alpha,\beta] \\ 0 & {\rm otherwise} \end{array} \right.
\end{equation}
The corresponding expression for $\Ts^{(1)}_{R}$ follows by replacing $[\alpha(0),\beta(0)]$ with $[\alpha(1),\beta(1)]$ in \eref{edgecorrectionfinalapp}. Finally we note that if $B_0(0)=0$ ($B_0(1)=0$) then $\Ts^{(1)}_{L}$ ($\Ts^{(1)}_{R}$) is zero. In particular, if the sequence is closed there are no edge contributions to $\Ts^{(1)}$.

\section{$\Ds^{(0)}$ and $\Ds^{(1)}$ for the Laguerre polynomials}
\label{appendixlaguerre}
Defining $A=(4\pi)^{-1}\sqrt{8 \lambda -\left(\lambda -\alpha _0\right)^2}$,
\begin{equation}
	\fl B=\frac{1}{\pi}\arccos\left(\frac{\alpha _0-\lambda+4}{2\sqrt{2\alpha_0+4}}\right)\mt{and}{1}C=\frac{1}{2\pi}\arccos\left(\frac{\alpha _0+\lambda }{2 \sqrt{\left(\alpha _0+2\right) \lambda }}\right)
\end{equation}
we find that $\Ds^{(0)}(\lambda)=A+B-\alpha_0 C$ and $\Ds^{(1)}(\lambda)=-A/2+B/4+(\alpha_0-2\alpha_1)C/2-1/8$.

\section*{References}
\bibliographystyle{plain}

\end{document}